\documentclass[reprint,
nofootinbib,
 amsmath,amssymb,
 aps,
pra,
]{revtex4-2}

\usepackage[utf8]{inputenc} 
\usepackage[T1]{fontenc}    
\usepackage{hyperref}       
\usepackage{url}            
\usepackage{booktabs}       
\usepackage{multirow}
\usepackage{amsfonts}       
\usepackage{nicefrac}       
\usepackage{microtype}      
\usepackage{graphicx}
\usepackage{xcolor, etoolbox}
\usepackage{amsmath,amsfonts,amsthm} 
\usepackage{mathtools}
\usepackage{ragged2e}
\usepackage{braket}      
\usepackage[english]{babel} 
\graphicspath{{./figures/}} 
\usepackage{enumitem}
\usepackage{xspace}
\usepackage{dsfont}
\usepackage{bm}
\usepackage{dcolumn}
\usepackage{mathrsfs}
\usepackage{arydshln}
\usepackage{soul}
\usepackage{placeins} 

\newcommand{\Tr}{\textbf{Tr}}

\hypersetup{
    colorlinks,
    linkcolor=[rgb]{0.206756, 0.371758, 0.553117},
    urlcolor=[rgb]{0.120565, 0.596422, 0.543611},
    citecolor=[rgb]{0.8784313725490196, 0.403921568627451, 0.4549019607843137},
    breaklinks=true 
}

\DeclarePairedDelimiter\abs{\lvert}{\rvert}
\newcommand{\norm}[1]{\left\lVert#1\right\rVert}

\DeclarePairedDelimiter\floor{\lfloor}{\rfloor}

\begin{document}

\title{Detecting strongly non-Gaussian entanglement}

\author{Serge Deside$^*$}
\author{Tobias Haas$^*$}
\author{Nicolas J. Cerf}
\affiliation{Centre for Quantum Information and Communication, École polytechnique de Bruxelles, CP 165, Université libre de Bruxelles, 1050 Brussels, Belgium}

\begin{abstract}
Efficiently certifying non-Gaussian entanglement in continuous-variable quantum systems is a central challenge for advancing quantum information processing, photonic quantum computing, and metrology. Here, we put forward continuous-variable counterparts to the recently introduced entanglement criteria based on moments of the partially transposed state, together with simple readout schemes that require only a few replicas of the state, passive linear optics, and particle-number measurements. We develop a multicopy method that enables the detection of strongly non-Gaussian entanglement for various relevant state families overlooked by standard approaches, which include the entire class of NOON states. Further, it is robust against realistic experimental constraints (losses, noise, and finite statistics), which we demonstrate by extensive numerical simulations.
\end{abstract}

\maketitle
\def\thefootnote{*}\footnotetext{These authors contributed equally to this work.}\def\thefootnote{\arabic{footnote}}

\section{Introduction}
\vspace{-10pt}

Over the last half-century, characterizing quantum entanglement emerged as a collective quest for understanding a broad variety of quantum phenomena, ranging from Bell tests and non-locality~\cite{Einstein1935,Bell1965,Clauser1969,Aspect1982,Weihs1998,Christensen2013}, over area laws and thermalization in many-body systems~\cite{Calabrese2004,Casini2009,Eisert2010,Abanin2012, Islam2015,Kaufman2016,Haas2024b,Haas2024c,Haas2024d} to black-hole physics~\cite{Bombelli1986,Srednicki1993,Callan1994,Solodukhin2011}. With the advent of quantum information processing, entanglement is nowadays considered an indispensable resource~\cite{Vedral1997,Plenio2007,Chitambar2019} for quantum computing~\cite{Jozsa2003,Knill2001,Kok2007,Slussarenko2019,Moody2022,Romero2024}, simulation~\cite{Bloch2012,Georgescu2014,Altman2021,Haas2022b,Haas2022c,Haas2025a} and metrology~\cite{DAriano2001,Paris2009,Giovanetti2011,Pezze2018}. Central to applications is its efficient detection by demonstrating the violation of a necessary condition on the target state's separability~\cite{Horodecki2009,Guehne2009}. This task becomes notoriously difficult in the realm of continuous-variable (CV) quantum systems, characterized by infinite-dimensional Hilbert spaces~\cite{Braunstein2005b,cerf2007quantum,Weedbrook2012,Serafini2017} -- including, e.g., photonics~\cite{Dong2008,Schneeloch2019,Asavanant2019,Qin2019} and ultracold quantum gases~\cite{Gross2011,Strobel2014,Peise2015,Kunkel2018,Lange2018,Fadel2018}.

Over the past three decades, a plethora of entanglement criteria have been put forward, most of which follow the well-known Peres-Horodecki \textit{positive-partial-transpose} (PPT) criterion~\cite{Peres1996,Horodecki1996}. The idea here is to test for negativities in the state's partial transpose $\tilde{\boldsymbol{\rho}} = (\mathds{1} \otimes T_B) \boldsymbol{\rho}$ via some measurement statistics, which  witness the presence of entanglement.
For instance, constraints on the quadratures' second moments fully classify Gaussian entanglement~\cite{Duan2000,Simon2000,Werner2001,Mancini2002,Giovannetti2003,Lami2018}. More advanced methods utilize entropies~\cite{Walborn2009,Walborn2011,Saboia2011,Tasca2013,Schneeloch2018,Haas2021b,Haas2022a,Haas2023a,Haas2023b}, which flag weak forms of non-Gaussian entanglement in, e.g., superpositions of Gaussian or close-to-Gaussian wavefunctions, or employ higher-order moments~\cite{Shchukin2005,Miranowicz2006,Miranowicz2009}. However, these approaches are inapplicable in the \textit{strongly non-Gaussian} regime, i.e., far beyond the convex hull of Gaussian states: Entropic criteria become insensitive, while the readout of higher-order moments may require infeasible amounts of measurement settings.
Given the importance of strongly non-Gaussian states for various protocols~\cite{PhysRevLett.89.137903,PhysRevLett.89.137904,PhysRevA.66.032316,PhysRevLett.102.120501,Walschaers2021,Chabaud2023} -- consider, e.g., NOON states~\cite{Sanders1989} and their optimality for metrology~\cite{Dowling2008} -- it is a major open challenge to devise strong and efficient entanglement criteria in this regime.

Recently, novel types of conditions testing for PPT have been introduced for discrete-variable systems. They constrain \textit{PT moments}, that is, moments $p_n = \Tr \{ \tilde{\boldsymbol{\rho}}^n \}$ of the partial transpose $\tilde{\boldsymbol{\rho}}$~\cite{Elben2020,Neven2021,Yu2021}, using randomized measurements serving as the readout technique for platforms comprising $\sim$10 qubits. However, their extensions to continuous variables -- including efficient methods for accessing PT moments -- have remained elusive.

In this paper, we extend these PT-moment-based criteria to bipartitions of CV systems and put forward simple interferometric measurement schemes for the readout of low-order PT moments by employing the \textit{multicopy} method~\cite{Brun2004}. Similarly as in other CV applications of this method~\cite{Hertz2019b,Arnhem2022,Griffet2023a,Griffet2023b}, our technique only requires passive optical elements acting on a few copies of the state and particle-number-resolving detectors, which have been established for both photonic~\cite{Marsili2013,Reddy2020,Cheng2023} and ultracold-atom~\cite{Bakr2009,Sherson2010,Hume2013,Gross2021} setups. We show that our method enables the measurement of PT moments of density matrices and is especially suited for detecting strongly non-Gaussian entanglement in both pure and mixed states, while remaining robust against common experimental imperfections.

\textit{The remainder of this paper is organized as follows.} In Sec.~\ref{sec:Criteria}, we recall and generalize the two most prominent hierarchies of PT-moment-based entanglement criteria (Sec.~\ref{subsec:pnCriteria}), with a particular focus on those involving PT moments up to order 3 (Sec.~\ref{subsec:ThirdOrderCriteria}). Then, in Sec.~\ref{sec:Implementation}, we devise measurement schemes based on the multicopy method (Sec.~\ref{subsec:MulticopyMethod}) to access PT moments  (Sec.~\ref{subsec:MeasuringPTMoments}) and we work out their linear-optics implementations for PT moments of order 2 and 3 (Sec.~\ref{subsec:OpticalCircuits}). The versatile detection capabilities of PT-moment-based criteria are benchmarked in Sec.~\ref{sec:Benchmarks} against standard methods (Sec.~\ref{subsec:Comparison}), for Gaussian states (Sec.~\ref{subsec:GaussianStates}), coherent superpositions/mixtures of Gaussian states (Sec.~\ref{subsec:NonGaussianStatesCoherent}), and strongly non-Gaussian states in Fock space (Sec.~\ref{subsec:NonGaussianStatesFock}). In Sec.~\ref{sec:Application}, we analyze the impact of realistic experimental conditions such as losses, noise, and finite statistics (Sec.~\ref{subsec:Imperfections}), and use simulated data to run a test case (Sec.~\ref{subsec:Simulation}). We discuss our findings and conclude in Sec.~\ref{sec:Discussion}.

\textit{Notation.} We employ natural units $\hbar =  1$ and use bold (normal) letters for quantum operators $\boldsymbol{O}$ (classical variables $O$). Single- and multi-copy \mbox{expectation} values are denoted by $\braket{\boldsymbol{O}} = \Tr \{ \boldsymbol{\rho} \, \boldsymbol{O} \}$ and $\braket{\cdots \braket{\boldsymbol{O}} \cdots} = \Tr \{ (\boldsymbol{\rho} \otimes \cdots \otimes \boldsymbol{\rho}) \, \boldsymbol{O} \}$, respectively. We associate the mode operators $\boldsymbol{a}$ and $\boldsymbol{b}$ with Alice's and Bob's subsystems $A$ and $B$, respectively. Further, we use a tilde for partially transposed quantities, e.g., $\tilde{\boldsymbol{\rho}}$, with the only exception being the PT moments noted as $p_n$.

\section{PT-moment-based criteria for continuous variables}
\label{sec:Criteria}
We start with extending the $p_n$-PPT criteria reported in~\cite{Elben2020,Neven2021,Yu2021} to continuous variables, with an emphasis on criteria involving the first two non-trivial PT moments.

\subsection{Infinite hierarchies of criteria}
\label{subsec:pnCriteria}

\subsubsection{From the Stieltjes moment problem}
\label{subsubsec:YuCriteria}
The first set of criteria is motivated by the question of whether a given set of PT moments $(p_1,\dots,p_n)$, defined via $p_n = \Tr \{ \tilde{\boldsymbol{\rho}}^n \}$, is compatible with a positive semi-definite partially transposed state $\tilde{\boldsymbol{\rho}} = (\mathds{1} \otimes T_B) \boldsymbol{\rho}$~\cite{Yu2021,Neven2021}. Corresponding to a quantum version of the classical Stieltjes moment problem, compatibility can be certified upon inspecting positivity conditions formulated in terms of Hankel matrices. Given the Hankel matrix over PT moments of increasing order along rows and columns
\begin{equation}
    P_n = \begin{pmatrix}
        p_1 & p_2 & p_3 & \cdots & p_{\frac{n+1}{2}}\\
        p_2 & p_3 & p_4 & \cdots & \\
        p_3 & p_4 & p_5 & \cdots  & \\
        \vdots & \vdots  & \vdots & \ddots & \\
        p_{\frac{n+1}{2}} & & & & p_n
    \end{pmatrix}, \quad n = 1, 3, 5, \dots,
\end{equation}
the $p_n$-PPT criterion states~\cite{Yu2021,Neven2021}
\begin{equation}
    \boldsymbol{\rho} \text{ separable} \implies P_n \ge 0, \quad \forall \text{ odd } n,
    \label{eq:pnPPTCriteria}
\end{equation}
such that violation of the latter inequality for some $n$ flags entanglement. Note that $p_1=1$ and $p_2$ equals the state's purity $\Tr \{ \boldsymbol{\rho}^2 \}$, so we need to exploit at least the third-order PT moment $p_3$ to flag entanglement. Sylvester's criterion implies $P_{n+2} \ge 0 \implies P_n \ge 0$, showing that higher-order criteria are stronger (and larger) sets of conditions on the positivity of $\tilde{\boldsymbol{\rho}}$. In fact,~\eqref{eq:pnPPTCriteria} is necessary and sufficient for PPT when considering \textit{all} PT moments (see Lemma 4 in~\cite{Yu2021}). 

Although the set of criteria~\eqref{eq:pnPPTCriteria} has been benchmarked and refined solely for discrete-variable systems, it readily extends to continuous variables since the Stieltjes-type moment problem is generally defined over non-negative numbers and, thus, does not constrain the dimension of the spectrum of the state. 

\subsubsection{From Descartes' rule of signs}
The second set of criteria tests for PPT by linking the eigenvalues of $\tilde{\boldsymbol{\rho}}$, more precisely, their elementary symmetric polynomials, to PT moments via Newton's identities~\cite{Neven2021}. The latter read
\begin{equation}
    n e_n = \sum_{j=1}^n (-1)^{j-1}e_{n-j}\ p_j, \quad n = 1,2, \dots,
    \label{eq:NewtonIdentities}
\end{equation}
where $e_n$ denotes the $n$th elementary symmetric polynomial evaluated at the eigenvalues $\{\lambda_1, \dots \}$ of $\tilde{\boldsymbol{\rho}}$, defined as
\begin{equation}
    e_n (\lambda_1, \dots) = \sum_{1 \leq j_1 < \cdots < j_n} \lambda_{j_1}\cdots\lambda_{j_n},
    \label{eq:ElementarySymmetricPolynomials}
\end{equation}
with $e_0(\lambda_1, \dots) = e_1(\lambda_1, \dots) = 1$. Then, 
\begin{equation}
    \boldsymbol{\rho} \text{ separable} \implies e_n \ge 0, \quad \forall \, n \in \mathbb{N},
    \label{eq:pnPPTCriteria2}
\end{equation}
from which conditions on the PT moments can be obtained via~\eqref{eq:NewtonIdentities}. Analogously to the criteria~\eqref{eq:pnPPTCriteria}, as a consequence of Descartes' rule of signs~\cite{Neven2021}, the set of conditions~\eqref{eq:pnPPTCriteria2} becomes equivalent to the non-negativity of $\tilde{\boldsymbol{\rho}}$ when $n \to \infty$~(see \cite{Curtiss1918}).

While the criteria~\eqref{eq:pnPPTCriteria2} have been formulated only for finite-dimensional Hilbert spaces, their extension to continuous variables is straightforward: Newton's identities hold equally for elementary symmetric functions, which formally generalize elementary symmetric polynomials in the limit of (countably) infinitely many eigenvalues, and so does Descartes' rule of signs when applied to the characteristic polynomial of an infinite-dimensional matrix.

\subsection{Third-order criteria}
\label{subsec:ThirdOrderCriteria}
Albeit the apparent outperformance of higher-order criteria, PT moments decrease exponentially when increasing $n$, thereby complicating entanglement detection in practice. This follows from $\abs{p_n} \le (p_2)^{n/2}$ for $n \ge 2$, which shows that higher-order PT moments are exponentially suppressed, especially if the considered state is highly mixed. On top of that, our proposed measurement scheme for $p_n$ requires $n$ replica states, which further increases the experimental burden for large $n$ (see Sec.~\ref{sec:Implementation}). 

Hence, PT-moment-based entanglement criteria limited to third order, that is, conditions involving only the purity $p_2$ and moment $p_3$, are of particular relevance for experimental applications. They are obtained from~\eqref{eq:pnPPTCriteria} and~\eqref{eq:pnPPTCriteria2} by setting $n=3$, leading to what we refer to as the \textit{quadratic}~\cite{Elben2020} and \textit{linear}~\cite{Neven2021} (in $p_2$) $p_3$-PPT criteria,
\begin{equation}
    \boldsymbol{\rho} \text{ separable} \implies p_3 - p_2^2 \ge 0, \quad p_3 - \frac{3 p_2 - 1}{2} \ge 0,
    \label{eq:p3PPTCriteria}
\end{equation}
respectively. Here, we have used  (for the quadratic criterion) that the positivity of a $2 \times 2$ matrix with positive elements is equivalent to the positivity of its determinant, and  (for the linear criterion) the relations $3 e_3 = e_2 - p_2 + p_3$ and  $ 2 e_2 = 1 - p_2$. Interestingly, the two conditions~\eqref{eq:p3PPTCriteria} coincide for pure states ($p_2=1$), giving $p_3\ge 1$, but also for mixed states with purity $p_2 = 1/2$, giving $p_3\ge 1/4$. In between, the quadratic criterion is stronger than the linear criterion for $p_2 < 1/2$, and vice versa when $1/2 < p_2 < 1$.

This observation suggests the existence of an \textit{optimal} criterion of third order, in the sense that given only the two lowest-order moments $p_2$ and $p_3$, no stronger condition on the separability of a state can be found. This is achieved by minimizing $p_3$ with respect to $p_2$ over separable states, which results in the optimal $p_3$-PPT criterion~\cite{Yu2021,Neven2021}
\begin{equation}
    \boldsymbol{\rho} \text{ separable} \implies p_3 - \alpha u^3 - (1 - \alpha u )^3 \ge 0,
    \label{eq:Optimalp3PPTCriterion}
\end{equation}
where $\alpha = \floor{1/p_2}$ denotes the floor function of $1/p_2$ and $u=\frac{\alpha + \sqrt{\alpha \left[p_2 (\alpha + 1) - 1 \right]}}{\alpha (\alpha + 1)}$ is a non-negative number. It is instructive to evaluate~\eqref{eq:Optimalp3PPTCriterion} when $1/2 < p_2 \leq 1$, in which case $\alpha = 1, u = (1 + \sqrt{2 p_2 - 1})/2$, such that~\eqref{eq:Optimalp3PPTCriterion} reduces to the linear criterion. Thus, the linear criterion is optimal for states of purity $1/2 \leq p_2 \leq 1$. In contrast, both the quadratic and linear criteria \eqref{eq:p3PPTCriteria} are outperformed by~\eqref{eq:Optimalp3PPTCriterion} for mixed states with $p_2 < 1/2$ (see Fig.~2 in~\cite{Neven2021}).  As already noted in~\cite{Yu2021}, the optimal $p_3$-PPT criterion~\eqref{eq:Optimalp3PPTCriterion} is inherently independent of the Hilbert space dimension, so it also constitutes a valid entanglement criterion in the CV regime. 

Remarkably, the third-order criteria are both necessary and sufficient conditions for pure-state entanglement -- also in the CV case. This follows directly from the Schmidt decomposition $\ket{\psi} = \sum_i c_i \ket{i,i}$, where $c_i \in \mathbb{C}$ are the Schmidt coefficients such that $\lambda_i = \abs{c_i}^2$ defines a probability distribution over a countably infinite set. In this case, PT moments reduce to $L^p$ norms, i.e., $p_n = \norm{\lambda}_n^n$ ($p_n = \norm{\lambda}_{n/2}^n$) for odd (even) $n$, which implies $p_n \le 1$. Equality for $n \ge 3$ holds if and only if all but one Schmidt coefficients vanish. Hence, pure product states are characterized by all PT moments being unity, while for pure entangled states, we have $p_n < 1$ for $n\ge 3$. 
 
\section{Implementation}
\label{sec:Implementation}
We now devise measurement schemes to access PT moments in CV systems. In particular, we provide passive linear optical circuits to be applied on replicas of the state, from which the two lowest-order moments, $p_2$ and $p_3$, can be extracted using particle-number measurements. 

\subsection{Multicopy method}
\label{subsec:MulticopyMethod}

On discrete-variable platforms like spin lattices, the PT moments are conveniently estimated from randomized measurements~\cite{Elben2020,Neven2021}. However, the efficiency of this ansatz -- in terms of the required measurements to extract a given PT moment with high accuracy -- relies \textit{heavily} on the Hilbert space dimension of a qubit being small. More precisely, the required number of measurements to ensure $\epsilon$ accuracy for an estimate of $p_n$ is on the order of $d^{\abs{AB}}/\epsilon^2$~\cite{Elben2020}, where $d$ denotes the single-spin Hilbert space dimension and $\abs{AB}$ the total number of spins of the bipartite system. Therefore, the randomized measurement toolbox breaks down in the CV regime. 

To overcome this limitation, we propose a readout scheme using the multicopy method~\cite{Brun2004} (see also~\cite{Ekert2002,Alves2003}). This technique has been successfully applied to entanglement witnessing in various bosonic systems~\cite{Daley2012,Islam2015,Kaufman2016,Gray2018,Griffet2023b}. It relies on the central observation that any $n$th order polynomial of the state $\boldsymbol{\rho}$ is measurable via a suitably chosen multicopy observable $\boldsymbol{O}_n$ defined over $n$ replicas of the state, denoted by $\boldsymbol{\rho}^{\otimes n}$. For CV optical systems, this observable can then be simply accessed by transforming it to particle-number measurements using passive, i.e., particle-number-preserving, transformations~\cite{Hertz2019b,Arnhem2022,Griffet2023a,Griffet2023b}. Counting particles is well established experimentally and is a standard requirement for photonic quantum computing~\cite{Knill2001,Kok2007,Slussarenko2019,Romero2024}, integrated circuits~\cite{Moody2022}, quantum gas microscopes \cite{Bakr2009,Sherson2010,Hume2013,Gross2021}, and when the system under scrutiny lacks a phase-stable local oscillator, e.g., in High Harmonic Generation (HHG)~\cite{Fuchs2022}. 

Given a multicopy observable $\boldsymbol{O}_n$, the remaining task is to find a passive unitary transformation $U$ appropriate for measuring $\boldsymbol{O}_n$ by counting particles at the output, 
\begin{equation}
    (\boldsymbol{a}_1, \dots, \boldsymbol{a}_n)^T \to U (\boldsymbol{a}_1, \dots, \boldsymbol{a}_n)^T,
\end{equation}
which can be efficiently decomposed into beam splitters
    \begin{equation}
    B_{\mu \nu} (\tau) = \begin{pmatrix}
        \sqrt{\tau} & \sqrt{1-\tau}\\
        \sqrt{1-\tau} & -\sqrt{\tau}
    \end{pmatrix},
\end{equation}
of transmissivity $\tau \in [0,1]$, and phase shifts
\begin{equation}
    R_{\mu} (\varphi) = e^{-i \varphi},
\end{equation}
with phase $\varphi \in [0, 2 \pi)$, acting on modes $\mu, \nu = 1, \dots, n$.

\subsection{Measuring PT moments}
\label{subsec:MeasuringPTMoments}
The required multicopy observable for accessing the $n$th PT moment $p_n$ on the $n$-copy state $\boldsymbol{\rho}^{\otimes n}$ reads~\cite{Elben2020}
\begin{equation}
    \begin{split}
        p_n &= \Tr \left\{ \boldsymbol{\rho}^{\otimes n} \left(\overrightarrow{\boldsymbol{\Pi}}_n^{A} \otimes \overleftarrow{\boldsymbol{\Pi}}_n^{B} \right) \right\} \\
        &\equiv \braket{ \cdots \braket{\overrightarrow{\boldsymbol{\Pi}}_n^{A} \otimes \overleftarrow{\boldsymbol{\Pi}}_n^{B}} \cdots },
    \end{split}
    \label{eq:MulticopyObservable}
\end{equation}
where $\overrightarrow{\boldsymbol{\Pi}}_n^{A}$ and $\overleftarrow{\boldsymbol{\Pi}}_n^{B}$ are normal, unitary (but not Hermitian\footnote{Although not Hermitian in general, the operators are normal and can therefore be treated as observables~\cite{Hu2017}. It is straightforward to verify that the multicopy method extends to normal operators.} for $n>2$) operators corresponding to cyclic shifts, i.e., permutations, of the Fock basis elements to the right on Alice's side and to the left on Bob's side, that is, 
\begin{equation}
    \begin{split}
        \overrightarrow{\boldsymbol{\Pi}}_n^{A} \ket{i_1, \dots, i_n}_A &= \ket{i_n, \dots, i_{n-1}}_A, \\
        \overleftarrow{\boldsymbol{\Pi}}_n^{B} \ket{i_1, \dots, i_n}_B &= \ket{i_2, \dots, i_1}_B,
    \end{split}
\end{equation}
respectively. Importantly, to evaluate the PT moment $p_n$, it suffices that Alice and Bob measure $\overrightarrow{\boldsymbol{\Pi}}_n^{A}$ and $\overleftarrow{\boldsymbol{\Pi}}_n^{B}$ \textit{separately} on their respective subsystems of the $n$-copy state, then evaluate the product of their outcomes (by communicating classically), and finally average over sufficiently many experimental runs.

To devise a suitable multicopy measurement scheme, let us first consider Alice's side. We introduce the discrete Fourier transform (DFT)
\begin{equation}
    \text{F} (n) = \frac{1}{\sqrt{n}} \begin{pmatrix}
        1 & 1 & \cdots & 1\\
        1 & \omega_n & \cdots & \omega^{n-1}_n\\
        \vdots & \vdots & \ddots & \vdots \\
        1 & \omega^{n-1}_n & \cdots & \omega^{(n-1) (n-1)}_n
    \end{pmatrix},
    \label{eq:dftN}
\end{equation}
with complex phases $\omega_n = e^{-2\pi i/n}$ and rank $n$.
\begin{figure*}
    \centering
    \includegraphics[width=0.92\linewidth]{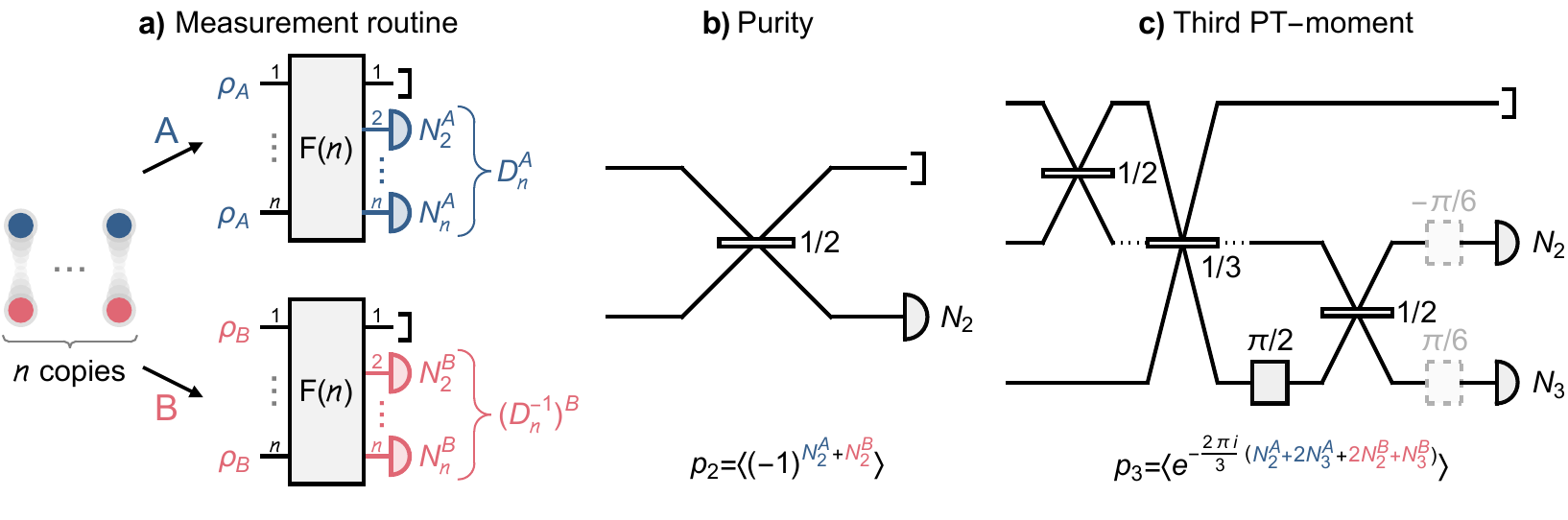}
    \caption{\textbf{a)} Measurement routine for detecting the $n$th PT moment $p_n$. After preparing $n$ replicas, that is, independent and identical copies, of the bipartite state of interest, Alice and Bob separately assess their subsystems $A$ (blue) and $B$ (red), respectively. Both perform a discrete Fourier transform (DFT) F$(n)$ of order $n$, see~\eqref{eq:dftN}, and then measure the particle numbers on their output modes $2$ to $n$. They obtain $p_n$ from the multicopy observable $\boldsymbol{D}_n^A \otimes (\boldsymbol{D}_n^{-1})^B$, see~\eqref{eq:DMatrix}. \textbf{b)} Readout of the purity $p_2$. The second-order DFT F$(2)$ is realized by a balanced beam splitter, upon which both Alice and Bob just measure particle numbers. \textbf{c)} Measurement of the third PT moment $p_3$. The third-order DFT F$(3)$ is implemented by a sequence of three beam splitters and three phase shifts (only the first is needed in practice), upon which incident particles are counted in the output modes $2$ and $3$. Here, Bob's multicopy observable corresponds to the inverse of Alice's, which is reflected in the particle numbers corresponding to the output modes $2$ and $3$ being interchanged for Bob.}
    \label{fig:Implementation}
\end{figure*}
It diagonalizes the right-shift operator 
$\overrightarrow{\boldsymbol{\Pi}}_n$ since the latter is circulant, so that we have
\begin{equation}
\overrightarrow{\boldsymbol{\Pi}}_n = \text{F}^\dagger (n) \boldsymbol{D}_n \text{F} (n),
\end{equation}
with the diagonal and unitary matrix
\begin{equation}
    \boldsymbol{D}_n = \omega_n^{\sum_{j=1}^n(j-1)\boldsymbol{N}_j} \equiv \text{diag} (1, \omega_n, \dots, \omega^{n-1}_n).
    \label{eq:DMatrix}
\end{equation}
As desired, the latter only involves the particle number operators $\boldsymbol{N}_j = \boldsymbol{a}_j^{\dagger} \boldsymbol{a}_j$ and hence requires nothing but particle-number measurements. Therefore, by acting with the DFT onto state $\boldsymbol{\rho}^{\otimes n}$, the measurement of the shift operator $\overrightarrow{\boldsymbol{\Pi}}_n$ translates into applying the optical circuit realizing the DFT to the target state $\boldsymbol{\rho}^{\otimes n} \to \text{F} (n) \boldsymbol{\rho}^{\otimes n}\text{F}^\dagger (n)$ followed by particle number measurements. Specifically, we have
\begin{equation}
\begin{split}
  \Tr \left\{ \boldsymbol{\rho}^{\otimes n} \overrightarrow{\boldsymbol{\Pi}}_n  \right\} &= \Tr \left\{ \left[ \text{F} (n) \boldsymbol{\rho}^{\otimes n} \text{F}^\dagger (n) \right] \boldsymbol{D}_n  \right\}  \\
  &= \braket{\cdots \braket{ \boldsymbol{D}_n}\cdots} ,
\end{split}
\end{equation}
where the multicopy expectation value is now taken on the $n$-mode output state of the DFT.


This procedure can easily be extended to the bipartite case, where Bob measures the left-shift operator $\overleftarrow{\boldsymbol{\Pi}}_n$.  Using the fact that the latter is the inverse of the right-shift operator together with the unitarity of the Fourier transform, we find 
\begin{equation}
    \overleftarrow{\boldsymbol{\Pi}}_n = \text{F}^{\dagger} (n) \boldsymbol{D}_n^{-1} \text{F} (n).
\end{equation}
Therefore, for the readout of the $n$th PT moment, Alice and Bob need to perform the \textit{same} $n$th-order DFT on their respective sides. This results in the state transformation $\boldsymbol{\rho}^{\otimes n} \to \left[\text{F} (n) \otimes \text{F} (n) \right] \boldsymbol{\rho}^{\otimes n} \left[ \text{F}^{\dagger} (n) \otimes \text{F}^{\dagger} (n) \right]$ and the subsequent measurements [see Fig.~\hyperref[fig:Implementation]{1a)}]
\begin{equation}
    p_n = \braket{\cdots \braket{\boldsymbol{D}_n^A \otimes (\boldsymbol{D}_n^{-1})^B} \cdots}.
\end{equation}
We note here that this implementation is reminiscent of the method put forward in~\cite{Daley2012} to access Rényi entropies, i.e., moments of the state $\boldsymbol{\rho}$ itself. The key difference in our case -- the partial transpose $\tilde{\boldsymbol{\rho}}$ -- results in Bob having to measure the \textit{inverse} of Alice's observable.

Alternatively, one may use the identity $\overleftarrow{\boldsymbol{\Pi}}_n = \text{F} (n) \boldsymbol{D}_n \text{F}^{\dagger} (n)$, showing that the left shift operator can also be diagonalized by applying an \textit{inverse} Fourier transform. In this case, Alice and Bob would have to measure precisely the same observables, but Bob would have to reverse the optical circuit Alice used.

\subsection{Optical circuits for low-order moments}
\label{subsec:OpticalCircuits}
It is left to decompose the DFT \eqref{eq:dftN} into passive optical elements. In general, any $n$-dimensional Gaussian unitary can be realized using at most $n (n-1)/2$ beam splitters (each possibly followed by a phase shift), which stems from the decomposition of the $n$-dimensional unitary group $U(n)$ into $U(2)$ subgroups~\cite{Reck1994,PhysRevA.57.R1477,Clements2016}. In the following, we explicitly devise optical circuits implementing F$(2)$ and F$(3)$, enabling the measurements of $p_2$ and $p_3$, respectively.

\subsubsection{Purity $p_2$}
\label{subsubsec:p2}
The protocol for obtaining the state's purity using two copies is arguably the most prominent application of the multicopy method~\cite{Brun2004,Daley2012,Arnhem2022,Griffet2023a} and has been successfully implemented in cold-atom systems~\cite{Islam2015,Kaufman2016} and integrated photonic devices~\cite{Goldberg2023}. It is well known that the two-mode DFT is nothing but a 50:50 beam splitter operation, i.e.,
\begin{equation}
    \text{F} (2) = \frac{1}{\sqrt{2}} \begin{pmatrix}
        1 & 1\\
        1 & -1
    \end{pmatrix}.
\end{equation}
Note that in this case, the DFT matrix is involutory; i.e., $\text{F}^{-1}(2)=\text{F}(2)$ since a shift in any direction corresponds to the swap operation, which is another way of seeing that $p_2$ equals the purity. Hence, a single beam splitter on each side suffices for Alice and Bob to access $p_2$.

The multicopy observable after the beam splitter has been applied reads
\begin{equation}
    p_2 = \braket{\braket{\omega_2^{\boldsymbol{N}_2^A} \otimes \omega_2^{\boldsymbol{N}_2^B}}} = \braket{(-1)^{N_2^A + N_2^B}}
\end{equation}
and relates to a parity measurement of the total particle number across the second modes, specifically, the output modes displaying destructive interference. Thus, the experimental procedure consists of Alice and Bob counting the number of particles in these modes, sharing their information via classical communication, and assigning the values $+ 1$ or $-1$ for an even or odd \textit{total} number of particles, respectively, and thereupon averaging over sufficiently many experimental runs [see Fig.~\hyperref[fig:Implementation]{1b)}].

\subsubsection{Third-order PT moment $p_3$}
\label{subsubsec:p3}
Following \eqref{eq:dftN}, the third-order DFT evaluates to
\begin{equation}
    \text{F} (3) = \frac{1}{\sqrt{3}} \begin{pmatrix}
        1 & 1 & 1\\
        1 & e^{-i\frac{2\pi}{3}} & e^{i\frac{2\pi}{3}}\\
        1 & e^{i\frac{2\pi}{3}} & e^{-i\frac{2\pi}{3}}
    \end{pmatrix},
\end{equation}
whose implementation requires a combination of three beam splitters and phase shifts each, namely,
\begin{equation}
    \begin{split}
        \text{F} (3) &= R_3 \left(\frac{\pi}{6}\right) R_2 \left(-\frac{\pi}{6}\right) \text{B}_{23} \left(\frac{1}{2}\right) \\
        &\hspace{0.4cm}\times R_3 \left(\frac{\pi}{2}\right) \text{B}_{13} \left(\frac{2}{3}\right) \text{B}_{12} \left(\frac{1}{2}\right)
    \end{split}
    \label{eq:F3Circuit}
\end{equation}
[see Fig.~\hyperref[fig:Implementation]{1c)}]. However, as we conduct particle-number measurements at the output, the two last phase shifts of $\pm \pi/6$ (light gray) can be left out. Then, the third-order PT moment $p_3$ is encoded in the multicopy observable
\begin{equation}
    \begin{split}
        p_3 &= \braket{\braket{\braket{\omega_3^{\boldsymbol{N}_2^A + 2 \boldsymbol{N}_3^A } \otimes \omega_3^{2 \boldsymbol{N}_2^B + \boldsymbol{N}_3^B }}}} \\
        &= \braket{(e^{-2\pi i/3})^{ N_2^A + 2 N_3^A + 2 N_2^B + N_3^B}}.
    \end{split}
\end{equation}
Importantly, the prefactors of Alice's number operators appearing in the exponent are now interchanged for Bob's.

\section{Benchmarks}
\label{sec:Benchmarks}
Next, we benchmark the detection capabilities of the various third-order criteria discussed in Sec.~\ref{subsec:ThirdOrderCriteria}. We discuss their performance in the form of a general comparison to other strong criteria, upon which we explore a variety of typical state families. This includes Gaussian states and strongly (also referred to as quantum~\cite{Walschaers2021}) non-Gaussian states -- divided into superpositions of coherent or Fock states -- including, e.g., NOON states, whose entanglement is known to be hard to certify. 

\subsection{Comparison}
\label{subsec:Comparison}

\subsubsection{Shchukin-Vogel criteria}
\label{subsubsec:ShchukinVogel}
The well-known Shchukin-Vogel hierarchy constitutes another complete set of criteria testing for PPT~\cite{Shchukin2005,Miranowicz2006,Miranowicz2009}. In contrast to PT-moment-based criteria, they involve moments of mode operators. The central idea is that the partially transposed state $\tilde{\boldsymbol{\rho}}$ is non-negative if and only if $\Tr \{ \tilde{\boldsymbol{\rho}} \boldsymbol{f}^{\dagger} \boldsymbol{f}\} \ge 0$ for all normally ordered $\boldsymbol{f}$. It is convenient to expand $\boldsymbol{f} = \sum_{i,j,k,l} c_{i j k l} \, \boldsymbol{a}^{\dagger i} \boldsymbol{a}^{j} \boldsymbol{b}^{\dagger k} \boldsymbol{b}^{l}$ for some complex-valued $c$, leading to a reformulation in terms of the matrix $D_{qrst,ijkl} = \Tr \{ \boldsymbol{\rho} \, \boldsymbol{a}^{\dagger r} \boldsymbol{a}^{q} \boldsymbol{a}^{\dagger i} \boldsymbol{a}^{j} \boldsymbol{b}^{\dagger t} \boldsymbol{b}^{s} \boldsymbol{b}^{\dagger k} \boldsymbol{b}^{l} \}$, namely~\cite{Griffet2023b},
\begin{equation}
    \sum_{\substack{q,r,s,t \\ i,j,k,l}} c^*_{qrst} c_{ijkl} \, \tilde{D}_{qrst,ijkl} \ge 0,
    \label{eq:ShchukinVogelCriteria}
\end{equation}
for all $c$. This is, in fact, equivalent to all principal minors of $\tilde{D}$ being non-negative, as implied by Sylvester's criterion. A specific criterion is then obtained by choosing a certain $\boldsymbol{f}$ (or equivalently $c$), which corresponds to selecting some submatrix of $\tilde{D}$. For instance, setting $\boldsymbol{f} = c_1 + c_2 \boldsymbol{a} + c_3 \boldsymbol{b}$ leads to a second-order criterion that is necessary and sufficient for Gaussian entanglement.

It is natural to ask how~\eqref{eq:ShchukinVogelCriteria} relates to PT-moment-based criteria. When considered over \textit{all} $c$, the three \textit{infinite} hierarchies~\eqref{eq:ShchukinVogelCriteria},~\eqref{eq:pnPPTCriteria}, and~\eqref{eq:pnPPTCriteria2} are all equivalent to PPT. Still, their main conceptual difference is that while practically useful criteria of the Shchukin-Vogel hierarchy comprise only a few low-order mode-operator moments, PT moments can contain mode-operator moments up to the highest order of the underlying state.

To make a more precise statement, we derive a relation between the linear $p_3$-PPT criterion~\eqref{eq:p3PPTCriteria} and~\eqref{eq:ShchukinVogelCriteria}. Starting from the state's normally ordered expansion $\boldsymbol{\rho}=\sum_{i,j,k,l} \rho_{ijkl} \boldsymbol{a}^{\dagger i} \boldsymbol{a}^j \boldsymbol{b}^{\dagger k} \boldsymbol{b}^l$ for complex-valued $\rho$, normally ordering its powers allows to write any PT moment as
\begin{equation}
    p_n = \sum_{i,j,k,l} [\mathfrak{p}_{n-1}(\rho)]_{ijkl} \, \tilde{d}_{ijlk},
    \label{eq:PTMomentModeOperators}
\end{equation}
where $d$ denotes the matrix of mode-operator moments $d_{ijkl} = \Tr \{ \boldsymbol{\rho} \, \boldsymbol{a}^{\dagger i} \boldsymbol{a}^j \boldsymbol{b}^{\dagger k} \boldsymbol{b}^l \}$. Note here that $\tilde{d}_{ijlk} = d_{ijkl}$, and hence moments of the partially transposed state correspond to reordered moments of the state itself. Further, $\mathfrak{p}_{n}$, as defined in~\eqref{eq:PTMomentModeOperators}, are polynomials in $\rho$ of order $n$ that arise by repeatedly applying the bosonic commutation relations to ensure normal order in the trace. For the two lowest-order polynomials $\mathfrak{p}_1$ and $\mathfrak{p}_2$, which are required for $p_2$ and $p_3$, respectively, we obtain
\begin{equation}
    \begin{split}
        [\mathfrak{p}_1 (x)]_{i j k l} &= x, \\
        [\mathfrak{p}_2 (x)]_{i j k l} &= \sum_{q,r,s,t} \sum_{u=0}^{\min (q,s)} \sum_{v=0}^{\min (r,t)}  \, u! \, v! \\
        &\hspace{0.5cm} \times \binom{q}{u} \binom{s}{u} \binom{r}{v} \binom{t}{v} \\
        &\hspace{0.5cm} \times x_{i-s+u \, q \, r \, l-t+v } \, x_{s \, j-q+u \, k-r+v \, t}.
    \end{split}
\end{equation}

We find a similar decomposition for the Shchukin-Vogel criteria after normally ordering $\boldsymbol{f}^{\dagger} \boldsymbol{f}$, namely,
\begin{equation}
    \Tr \{ \tilde{\boldsymbol{\rho}} \boldsymbol{f}^{\dagger} \boldsymbol{f} \} = \sum_{i,j,k,l} [\mathfrak{f}_2 (c)]_{i j k l} \, \tilde{d}_{ijlk} \ge 0,
    \label{eq:ffdaggerModeOperators}
\end{equation}
where $\mathfrak{f}_2$ is defined in analogy to $\mathfrak{p}_2$, with the main difference being that indices are slightly reordered since $\mathfrak{f}_2$ stems from $\boldsymbol{f}^{\dagger} \boldsymbol{f}$, whereas $\mathfrak p_2$ stems from $\tilde{\boldsymbol{\rho}}^3$. In general, the map $c \to \mathfrak{f}_2 (c)$ is not surjective for complex-valued $c$. However, when identifying linear combinations of~\eqref{eq:PTMomentModeOperators} with~\eqref{eq:ffdaggerModeOperators}, $\mathfrak{f}_2 (c)$ must be Hermitian, in which case it can always be expressed as a bilinear form of some complex-valued $c$. 

Therefore, the linear $p_3$-PPT criterion~\eqref{eq:p3PPTCriteria} equals the Shchukin-Vogel criterion~\eqref{eq:ffdaggerModeOperators}, with the matrix $c$ being defined implicitly via $[\mathfrak{f}_2 (c)]_{i j k l} = [\mathfrak{p}_2 (\rho)]_{ijkl} - (3/2) \rho_{ijkl} + 1/2$. This shows that the set of mode-operator moments constrained by a PT-moment-based criterion depends solely on the state and can, in principle, range up to arbitrarily high orders -- a crucial property for capturing higher-order correlations (we provide a worked-out example for NOON states in Appendix~\ref{app:NOONMoments}). Given that the efficient readout of low-order Shchukin-Vogel criteria also requires multiple copies~\cite{Griffet2023b}, the herein proposed scheme is expected to outrange the Shchukin-Vogel method for strongly non-Gaussian states in terms of both detection capability and experimental resources (see, e.g., Sec.~\ref{subsubsec:NOONStates}).

\subsubsection{Criteria based on non-local variables}
\label{subsubsec:NonlocalVariables}
In another common approach, PPT is assessed via the non-local quadrature operators $\boldsymbol{x}_{\pm} = \boldsymbol{x}_A \pm \boldsymbol{x}_B$ and $\boldsymbol{p}_{\pm} = \boldsymbol{p}_A \pm \boldsymbol{p}_B$ from the EPR argument~\cite{Einstein1935}. By noting that taking the partial transpose in subsystem $B$ translates into $p_B \to - p_B$ in phase space~\cite{Simon2000}, physicality (meaning non-negativity) of $\tilde{\boldsymbol{\rho}}$ implies that uncertainty relations must be fulfilled by the mixed commuting variable pairs $(\boldsymbol{x}_{\pm}, \boldsymbol{p}_{\mp})$ for all separable states. 

Many entanglement criteria have been formulated in terms of the outcome distributions of a homodyne measurement $f_{\pm} (x_{\pm}) = \braket{x_{\pm} | \boldsymbol{\rho} | x_{\pm}}$, $g_{\mp} (p_{\mp}) = \braket{p_{\mp} | \boldsymbol{\rho} | p_{\mp}}$. This includes constraints on their second moments, such as $\sigma_{x_{\pm}}^2 + \sigma_{p_{\mp}}^2 \ge 2$~\cite{Duan2000} (see also~\cite{Simon2000,Werner2001,Mancini2002,Giovannetti2003,Lami2018}), and entropies, like $S(f_{\pm}) + S(g_{\mp}) \ge \ln (2 \pi e)$~\cite{Walborn2009} (see also~\cite{Walborn2011,Saboia2011,Tasca2013,Schneeloch2018}). Recently, entanglement criteria were introduced for the Husimi-$Q$ distribution $Q_{\pm} (x_{\pm}, p_{\mp})$ associated with a heterodyne measurement, ranging from entropic variants $S(Q_{\pm}) \ge \ln (2e)$~\cite{Haas2022a} (see also~\cite{Haas2021b}) to majorization relations $Q_{\pm} \prec Q_{\pm,\text{vacuum}}$~\cite{Haas2023a,Haas2023b}.

While all these methods fully characterize Gaussian entanglement, the entropic criteria also witness some forms of non-Gaussian entanglement, such as superpositions of Gaussian states, e.g., the entangled cat state (see Sec.~\ref{subsubsec:MixedEntangledCat}) and close-to-Gaussian states, e.g., the Hermite-Gauss state [see Eq.~\eqref{eq:HermiteGaussState}]. However, entanglement cannot be certified over the whole parameter range in both cases. Additionally, strongly non-Gaussian entanglement arising from finite superpositions in Fock space is entirely overlooked. Thus, such methods are efficient when dealing with Gaussian and Gaussian-like entanglement, but remain insensitive beyond. 

Intuitively, this is rooted in how (quantum) correlations are analyzed: The non-local variables $(x_{\pm},p_{\mp})$ are linear and hence capture mostly linear correlations between the two subsystems, while lacking information about higher-order correlations. Therefore, we expect our method to outrange all criteria based on EPR-type variables in the strongly non-Gaussian regime (see, e.g., Sec.~\ref{subsec:NonGaussianStatesFock}).

\subsection{Gaussian entanglement}
\label{subsec:GaussianStates}

\subsubsection{Symplectic formalism and Simon's criterion}
\label{subsubsec:SymplecticFormalism}
It is convenient to group the two pairs of canonical operators into a single vector in phase space as
\begin{equation}
    \boldsymbol{\chi} = (\boldsymbol{X}_1, \boldsymbol{P}_1, \boldsymbol{X}_2, \boldsymbol{P}_2)^T.
    \label{eq:PhaseSpaceOperator}
\end{equation}
Then, the canonical commutation relations take the form $\left[ \boldsymbol{\chi}_j, \boldsymbol{\chi}_{j'} \right] = i\Omega_{j j'}$, where $\Omega = \mathds{1}_2 \otimes (i\sigma_2)$ denotes the symplectic metric. A partially transposed Gaussian state is fully characterized by its mean vector and covariance matrix
\begin{equation}
    \tilde{\chi} = \Tr \left\{ \tilde{\boldsymbol{\rho}} \, \boldsymbol{\chi}\right\}, \quad \tilde{\gamma}_{j j'} = \Tr \left\{ \tilde{\boldsymbol{\rho}} \, \{ \boldsymbol{\chi}_j, \boldsymbol{\chi}_{j'} \} \right\} - 2\tilde{\chi}_j\tilde\chi_{j'},
    \label{eq:MeanAndCovarianceMatrix}
\end{equation}
respectively. While the former is irrelevant for separability, the latter is non-negative for all states but fulfills the stronger condition $\tilde{\gamma} + i\Omega \ge 0$ if and only if $\boldsymbol{\rho}$ is Gaussian and separable, which is known as Simon's criterion~\cite{Simon2000} (other well-known criteria have been introduced by Duan et al.~\cite{Duan2000} and MGVT~\cite{Mancini2002,Giovannetti2003}). It can be recast into a simple form in terms of the so-called symplectic eigenvalues of $\tilde{\gamma}$. To this end, we employ Williamson's theorem~\cite{Williamson1936}, which ensures the existence of a symplectic matrix $\mathcal{S}$ that diagonalizes the covariance matrix such that $\mathcal{S} \, \tilde{\gamma} \, \mathcal{S}^T = \text{diag} (\tilde{\nu}_1, \tilde{\nu}_1, \tilde{\nu}_2, \tilde{\nu}_2)$, where $\tilde{\nu}_1$ and $\tilde{\nu}_2$ denote the symplectic eigenvalues. Hence, a Gaussian state is separable if and only if
\begin{equation}
    \tilde{\nu}_1, \tilde{\nu}_2 \ge 1
    \label{eq:SimonCriterionSymplectic}
\end{equation}
[see the straight black curves in Figs.~\hyperref[fig:Benchmarks]{2a)} and~\hyperref[fig:Benchmarks]{2b)}].

\subsubsection{$p_n$-PPT criteria for Gaussian states}
\label{subsubsection:GaussianMoments}
A closed formula for the $n$th moment of a Gaussian state in terms of its symplectic eigenvalues has been derived in~\cite{Serafini2017,Seshadreesan2018}. For the partially transposed state, it reads
\begin{equation}
    p_n = \prod_{j=1}^2 \, \frac{2^{n}}{(\tilde{\nu}_j + 1)^n - (\tilde{\nu}_j - 1)^n}.
    \label{eq:GaussianMoments}
\end{equation}
For $n=2$ and $n=3$, the formula evaluates to
\begin{equation}
    p_2 = \frac{1}{\tilde{\nu}_1 \tilde{\nu}_2} = \frac{1}{\sqrt{\det \tilde{\gamma}}}, \quad p_3 = \frac{16}{\left(1 + 3 \tilde{\nu}_1^2 \right) \left(1 + 3 \tilde{\nu}_2^2 \right)},
    \label{eq:p2p3Gaussian}
\end{equation}
respectively. Hence, a Gaussian state is pure if and only if the product of symplectic eigenvalues equals unity, i.e., $\tilde{\nu}_1 \tilde{\nu}_2 = 1$. This implies that all states have to respect the physicality condition $\tilde{\nu}_1 \tilde{\nu}_2 \ge 1$ (see dashed black curves), which is equivalent to $\tilde{\gamma}$ being non-negative ($\tilde{\nu}_1 \tilde{\nu}_2 = \nu_1 \nu_2$ is ensured by symplectic matrices having unit determinant). 

With~\eqref{eq:p2p3Gaussian} at hand, we can express the $p_3$-PPT criteria in terms of the symplectic eigenvalues $(\tilde{\nu}_1, \tilde{\nu}_2)$ only. The quadratic $p_3$-PPT criterion reads (blue curves)
\begin{equation}
    7 \tilde{\nu}_1^2 \tilde{\nu}_2^2 - 3 \left( \tilde{\nu}_1^2 + \tilde{\nu}_2^2 \right) - 1 \ge 0,
\end{equation}
while the linear criterion evaluates to (green curves)
\begin{equation}
    (\tilde{\nu}_1^2+3\tilde{\nu}_1^2\tilde{\nu}_2^2+\tilde{\nu}_2^2)(\tilde{\nu}_1\tilde{\nu}_2-3) + 11\tilde{\nu}_1\tilde{\nu}_2 - 1 \ge 0.
\end{equation}
It is straightforward to check that both conditions and the optimal criterion (red curves) are implied by Simon's criterion~\eqref{eq:SimonCriterionSymplectic}, showing that all three are necessary but not sufficient conditions for Gaussian separability. Sufficiency is retained in the limit $n \to \infty$, see Appendix~\ref{app:GaussianStates}, in accordance with PPT being equivalent to separability for $1 \times 1$ Gaussian states~\cite{Werner2001}.

\subsubsection{Optimal PT-moment-based criterion for Gaussian entanglement}
\label{subsubsec:OptimalGaussianCriterion}
When expressing Simon's criterion~\eqref{eq:SimonCriterionSymplectic} through the two lowest-order PT moments $p_2$ and $p_3$, we find that a Gaussian state is separable if and only if
\begin{equation}
    p_3 \ge \frac{4 p_2^2}{3 + p_2^2},
    \label{eq:SimonCriterionMoments}
\end{equation}
which constitutes the optimal $p_3$-PPT criterion for Gaussian entanglement (see straight black curves). However, note that this criterion is biased in the sense that it only holds under the assumption that the underlying state is of Gaussian form. Translating also the physicality condition $\tilde{\nu}_1 \tilde{\nu_2} \ge 1$ to PT moments by using $\tilde{\nu_2} = 1/(\tilde{\nu_1} p_2)$ for $p_3$ and maximizing over $\tilde{\nu_1}$ yields an upper bound constraining all Gaussian states (see dashed black curves)
\begin{equation}
    p_3 \le \left(\frac{4 p_2}{3 + p_2}\right)^2.
\end{equation}
Of course, one might argue that requiring copies of an entangled Gaussian state to witness its entanglement is unnecessarily demanding; however, the method becomes powerful in the non-Gaussian regime.

\subsection{Strongly non-Gaussian entanglement from coherent states}
\label{subsec:NonGaussianStatesCoherent}

\subsubsection{Mixed entangled cat states}
\label{subsubsec:MixedEntangledCat}
Non-Gaussianity naturally arises from mixtures of Gaussian states, while coherent superpositions are required to generate strongly non-Gaussian states, for which Schrödinger cat states display an important example~\cite{Yurke1986,Walschaers2021}. In the following, we consider the dephased [even ($+$) or odd ($-$)] entangled Schrödinger cat states
\begin{equation}
    \begin{split}
        \boldsymbol{\rho} &= \frac{1}{C}\big[\ket{\alpha,\beta}\bra{\alpha,\beta}+\ket{-\alpha,-\beta}\bra{-\alpha,-\beta} \\ 
        &\hspace{0.35cm}\pm(1-z)\big(\ket{\alpha,\beta}\bra{-\alpha,-\beta}+\ket{-\alpha,-\beta}\bra{\alpha,\beta}\big)\big],
    \end{split}
    \label{eq:CatStates}
\end{equation}
with normalization $C = 2 \big[1\pm(1-z) e^{-2(|\alpha|^2+|\beta|^2)}\big]$, displacements $\alpha, \beta \in \mathbb{C}$, and mixing (dephasing) parameter\footnote{Centering both cats  ($\alpha = \beta = 0$), i.e., reducing the state to the vacuum, requires $z >0$ for the state to be well defined.} $z \in [0,1]$. The state's purity decreases monotonically\footnote{The only exception being when $\abs{\alpha}, \abs{\beta} \lesssim 1$ for the odd cat state.} for increasing $z$, with $p_2 = 1$ if and only if $z=0$, and $p_2 = 1/2$ for $z =1$ and $\abs{\alpha},\abs{\beta} \to \infty$. When $z=1$, $\alpha=0$, or $\beta = 0$, the state~\eqref{eq:CatStates} is separable, but entanglement is challenging to reveal for odd cat states: Second-moment-based criteria fail to witness entanglement, while entropic criteria flag entanglement only if the two cats are well separated, i.e., beyond $\abs{\alpha} = \abs{\beta} \gtrsim 3/2$~\cite{Walborn2009,Saboia2011,Haas2022a}.

We provide expressions for the second and third PT moments in Appendix~\ref{app:CatStates}. Since $p_2 \ge 1/2$ over the full parameter range, the linear criterion is optimal. Entanglement is always witnessed for even cat states (as in~\cite{Gessner2016}). For odd cat states, we obtain the separability condition
\begin{equation}
    \abs{\alpha}^2 + \abs{\beta}^2 \le - \frac{1}{2} \ln (1-z)
\end{equation}
[see Fig.~\hyperref[fig:Benchmarks]{2c)}]. For given $z$, entanglement is certified in the complement of a disk of radius $\sqrt{- (1/2) \ln (1-z)}$ in the $(\abs{\alpha}, \abs{\beta})$-plane. For $\abs{\alpha} = \abs{\beta}$, this amounts to $\abs{\alpha} \ge \sqrt{-\ln (1-z)}/2$ (see inset), which is smaller than $3/2$ up to $z = 1 - e^{-9} \approx 0.99988$. Hence, all entropic criteria are outperformed by the $p_3$-PPT criteria. Note also that entanglement is certified for all $\abs{\alpha}, \abs{\beta} > 0$ when the state is pure ($z=0$).

\subsubsection{High Harmonic Generation (HHG)}
Assessing entanglement by counting photons is particularly relevant for applications where other standard readout methods are inapplicable. A prime example of recent interest is the process of High Harmonic Generation (HHG), in which intense laser fields are directed at atomic media to generate high-order harmonics of the driving field~\cite{Franken1961}. In the strongly interacting regime, i.e., when the generated light lies in the extreme ultraviolet (XUV) spectrum, a phase-stable reference beam is often lacking, thereby preventing quadrature measurements via both the homodyne and heterodyne detection protocols~\cite{Fuchs2022}.

In a recently proposed multimode quantum description of HHG, the initial coherent-state driving field $\alpha$ (mode $1$) is depleted by $\delta \alpha > 0$, which causes displacements $\chi_j$ in the $N-1$ harmonic modes $j=2, \dots, N$, leading to entanglement between the driving laser and the higher harmonics, as well as among the harmonics themselves~\cite{Stammer2022}. The corresponding global state reads
\begin{equation}
    \begin{split}
    \ket{\psi}=\frac{1}{\sqrt{C}}\Big(&\ket{\alpha-\delta\alpha}\otimes_{j=2}^{N}\ket{\chi_j} \\
    &-\braket{\alpha|\alpha-\delta\alpha}\ket{\alpha}\otimes_{j=2}^{N}\braket{0_j|\chi_j}\ket{0_j} \Big),
    \end{split}
    \label{eq:hhg}
\end{equation}
with normalization $C = 1-e^{-|\delta\alpha|^2-\sum_{j=2}^{N} |\chi_j|^2}$. 

In the following, we consider all harmonic modes to be displaced equally, i.e., $\chi = \chi_j$, in which case ${\abs{\chi} \propto \abs{\delta \alpha}}$ by energy conservation and only odd harmonics are populated. We analyze entanglement across the bipartition of the driving field and one harmonic described by the mixed state 
\begin{equation}
    \boldsymbol{\rho} = \Tr_{j>2} \left\{ \ket{\psi} \bra{\psi} \right\};
    \label{eq:HHGState}
\end{equation}
see Appendix~\ref{app:HHGStates} for an explicit expression and the corresponding PT moments. 

\begin{figure*}
    \centering
\includegraphics[width=0.85\textwidth]{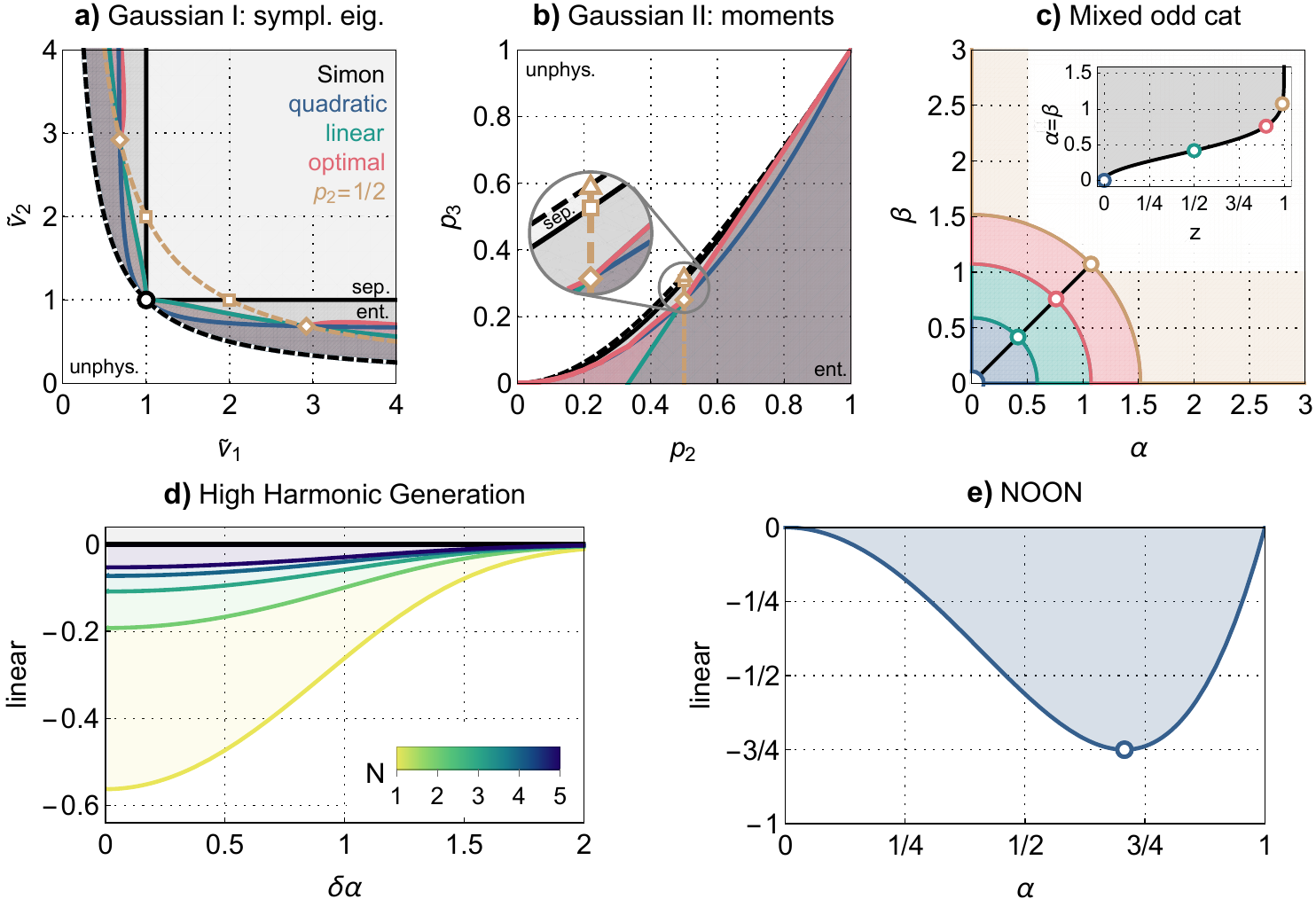}
    \caption{\textbf{a)} Detected Gaussian entanglement by PT-moment-based criteria of order 3 (shaded regions) in terms of the symplectic eigenvalues $\tilde\nu_1, \tilde\nu_2$. The set of physical Gaussian states constrained by $\tilde{\nu}_1 \tilde{\nu}_2 \ge 1$ (above and to the right of the black dashed line) is divided by Simon's criterion~\eqref{eq:SimonCriterionSymplectic} (black straight line) into separable ($\tilde\nu_1 \ge 1$ and $\tilde\nu_2 \ge 1$) and entangled ($\tilde\nu_1 < 1$ or $\tilde\nu_2 < 1$) regions. The performance across the criteria is governed by the purity $p_2$, with the tipping point at $p_2=1/2$ (yellow dashed line). This purity is realized, for instance, by a two-mode squeezed state at finite temperature, see Appendix~\ref{app:TMSState}, with $\bar{n} = (\sqrt{2}-1)/2$ particles per mode. No third-order criterion detects entanglement when $\nu_1 \lesssim 1, \nu_2 \gg 1$ (or vice versa). \textbf{b)} Complementary analysis in terms of the PT moments $p_2$ and $p_3$. \textbf{c)} Certified entanglement for the odd Schrödinger cat state~\eqref{eq:CatStates} by the optimal ($=$ linear since $p_2 \ge 1/2$) $p_3$-PPT criterion for various mixing parameters $z=0, 0.5, 0.9, 0.99$ (blue, green, red, and yellow, respectively) and (without loss of generality real) phases $\alpha, \beta$. Entanglement is flagged well below $\alpha = \beta \approx 3/2$ (see inset), thereby outperforming all entropic methods. \textbf{d)} Absolute violation of the linear $p_3$-PPT criterion for the entanglement between the driving field and the first harmonic mode in High Harmonic Generation (HHG), see~\eqref{eq:HHGState}, as a function of the (without loss of generality real) depletion $\delta \alpha$ of the laser field for various total numbers of excited harmonics $N$. Entanglement is witnessed over the full parameter range. \textbf{e)} The same as in panel d) for the family of NOON states~\eqref{eq:NOONState} as a function of the (without loss of generality real) phase $\alpha$. Entanglement is detected for all mode numbers $N$ and $\alpha \neq 0,1$, with the strongest violation $-3/4$ occurring at $\alpha = 1/\sqrt{2} \approx 0.707$, independent of $N$. All other standard methods are significantly outperformed.}
    \label{fig:Benchmarks}
\end{figure*}

Entanglement is detected by the linear $p_3$-PPT criterion (optimal because $p_2 \ge 1/2$ in the considered regime), also for many harmonic modes [see Fig.~\hyperref[fig:Benchmarks]{2d)}]. Note that the purity $p_2$ is monotonically increasing with $\delta \alpha$, and that $p_2 = 1$ for $N=1$, while $p_2 \gtrsim 4/5$ for all $N>1$. The amount of entanglement behaves very similarly when considered as a function of $\delta \alpha$ with a pure product state being attained in the limit $\delta \alpha \to \infty$, see also Fig.~3 in~\cite{Stammer2022}, and hence entanglement is harder to certify for more pure states. Further, we observe a decrease in absolute violation for increasing $N$, as the entanglement is split among more modes.

It was proposed in Ref.~\cite{Stammer2022} that entanglement can be enhanced by using conditioning measurements on complementary modes instead of the partial trace. In this case, the remaining bipartition is always pure, and thus entanglement is fully witnessed by any $p_3$-PPT criterion.

\subsection{Strongly non-Gaussian entanglement from Fock states}
\label{subsec:NonGaussianStatesFock}

\subsubsection{NOON states}
\label{subsubsec:NOONStates}
Entanglement detection is hardest when considering finite superpositions in Fock space. Of particular importance is the class of bosonic NOON states~\cite{Sanders1989}
\begin{equation}
    \ket\psi = \alpha \ket{N,0} + \beta \ket{0,N},
    \label{eq:NOONState}
\end{equation}
with mode population $N\in \mathbb{N}^+$ and phases $\alpha, \beta \in \mathbb{C}$ constrained by $\abs{\alpha}^2 + \abs{\beta}^2 = 1$ due to state normalization. This class comprises the first two Bell states for $N=1$ and $\alpha = 1/\sqrt{2}, \beta = \pm 1/\sqrt{2}$, and the Hong-Ou-Mandel state for $N=2$ and $\alpha = - \beta = 1/\sqrt{2}$. States of the form~\eqref{eq:NOONState} have successfully been prepared deterministically for $N \le 2$, e.g., using Hong-Ou-Mandel interferometry. In contrast, higher-order NOON states with $N > 2$ were generated probabilistically by additionally invoking post-selection (see~\cite{Walther2004,Mitchell2004,Resch2007,Nagata2007,Afek2010,Israel2012,Slussarenko2017}). Although NOON states are evidently entangled whenever neither of the complex phases is zero, detecting their entanglement is notoriously difficult -- both theoretically and experimentally. Non-local EPR-type criteria -- whether based on second moments or entropies -- never flag entanglement~\cite{Duan2000,Mancini2002,Giovannetti2003,Walborn2009,Walborn2011,Saboia2011,Tasca2013,Schneeloch2018,Haas2021b,Haas2022a,Haas2023a,Haas2023b}, while mode-operator-based criteria require moments up to order $2N$~\cite{Callus2025}, for which no practical readouts exist when $N$ becomes large. For instance, a recent criterion formulated for a spin-like multicopy observable encoding fourth-order moments witnesses entanglement only up to $N=2$~\cite{Griffet2023b}.

Since NOON states are pure ($p_2 = 1$), their entanglement is certified by all third-order criteria. More precisely, after a straightforward calculation, we find for $p_3$
\begin{equation}
    \begin{split}
        p_3 &= \Tr \Big\{ |\alpha|^6\ket{N,0}\bra{N,0} + \alpha \beta^* |\alpha|^2|\beta|^2\ket{N,N}\bra{0,0} \\ &\hspace{0.55cm}+\alpha^*\beta|\alpha|^2|\beta|^2\ket{0,0}\bra{N,N} + |\beta|^6\ket{0,N}\bra{0,N} \Big\} \\
        &= |\alpha|^6 + |\beta|^6.
    \end{split}
    \label{eq:p3NOONState}
\end{equation}
Remarkably, this result is independent of the mode population $N$ (which generalizes to all PT moments) and $p_3$ is strictly less than unity if and only if $\alpha$ or $\beta \neq 0$ (see Fig.~\hyperref[fig:Benchmarks]{2e)}). This demonstrates that PT-moment-based criteria significantly outperform all other approaches in the strongly non-Gaussian regime (see Appendix~\ref{app:NOONMoments} for an in-depth analysis of which mode-operator-moments appear in the PT moments). These findings also persist under realistic experimental constraints, which effectively render the underlying state mixed (see Sec.~\ref{subsec:Simulation}).

\subsubsection{Other pure-state families}
In the following, we list a few important pure entangled states that belong to the class of strongly non-Gaussian states, which are all automatically witnessed by any $p_3$-PPT criterion. Some of them are typical benchmarks for entanglement criteria, while others are of particular experimental relevance.

\begin{itemize}
    \item Entangled Fock-state superpositions
    \begin{equation}
        \ket{\psi} = \sum_{j, j'=0}^{N ,N'} c_{j j'} \ket{j,j'},
    \end{equation}
    which includes, e.g., the photonic qutrit state~\cite{Rodo2008,Walborn2009} $\ket{\psi} = \left( \sqrt{2} \ket{0,0} + \ket{2,0}+\ket{0,2} \right)/2$, whose entanglement remains undetected by entropic methods~\cite{Walborn2009,Walborn2011,Saboia2011,Tasca2013,Schneeloch2018,Haas2021b,Haas2022a,Haas2023a,Haas2023b}.

    \item Photon-added and photon-subtracted states, which are obtained by adding or subtracting photons from a pure Gaussian state, i.e.,
    \begin{equation}
        \hspace{0.75cm} \ket{\psi_+} \propto (\boldsymbol{a}^\dagger)^j\ket{\psi_{\text{Gauss}}}, \,\, \ket{\psi_-} \propto (\boldsymbol{a})^j\ket{\psi_{\text{Gauss}}},
    \end{equation}
    respectively~\cite{Walschaers2021}. Both processes enhance the entanglement in the initial Gaussian state. Therefore, entanglement is typically witnessed also by second-moment criteria as long as $j$ remains small~\cite{Duan2000,Simon2000,Werner2001,Mancini2002,Giovannetti2003,Lami2018}. 
    
    \item The Hermite-Gauss state described by the wavefunction~\cite{Rodo2008,Walborn2009}
    \begin{equation}
        \hspace{0.7cm} \psi(x_1,x_2)=\frac{x_1+x_2}{\sqrt{\pi\sigma_{-}\sigma_{+}^3}}e^{-\big(\frac{(x_1+x_2)^2}{4\sigma_{+}^2}+{\frac{(x_1-x_2)^2}{4\sigma_{-}^2}}\big)},
        \label{eq:HermiteGaussState}
    \end{equation}
    whose generation involves the non-Gaussian input $\ket{0,1}$ fed into a Gaussian circuit~\cite{Hertz2016}. No entropic criterion detects entanglement over the full parameter range $\sigma_+, \sigma_- > 0$~\cite{Walborn2009,Walborn2011,Saboia2011,Tasca2013,Schneeloch2018,Haas2021b,Haas2022a,Haas2023a,Haas2023b}.
\end{itemize}

\section{Application}
\label{sec:Application}

Robustness to experimental imperfections is an essential requirement of any practically useful entanglement criterion. First, we discuss typical sources of errors encountered in our detection schemes and how to model them. Then, we explore their influence in detail for NOON states. 

\subsection{Imperfections}
\label{subsec:Imperfections}
We are interested in the three main types of imperfections: the inherent losses that occur in state preparation, the optical circuit and detection; the noise-induced discrepancies between different state copies; and the limitations of finite measurement statistics.

\subsubsection{Losses}
\label{subsubsec:Losses}
Particle losses are inevitable in the preparation process of a given input state $\boldsymbol{\rho}$, the optical circuit, and the detection of single particles. With advances in superconducting photon-number-resolving detectors, efficiencies in photonic setups are now on the order of $\sim 95 \%$~\cite{Marsili2013,Reddy2020} and single photons can be resolved up to $100$ photons~\cite{Cheng2023} (see also~\cite{Slussarenko2019} and references therein). In ultracold atom experiments, quantum gas microscopes enable single-particle resolution up to $\sim 1000$ particles \cite{Bakr2009,Sherson2010,Hume2013} (see \cite{Gross2021} for a review). Therefore, losses are most detrimental during state preparation and optical processing.

The effect of losses is conveniently characterized by a lossy quantum channel $\mathcal{L}_\tau$, where $\tau$ is the loss parameter, being unity in the lossless case~\cite{Weedbrook2012}. Since losses can primarily be attributed to the optical circuit, we assume equal losses for both subsystems $A$ and $B$, as well as among the copies of $\boldsymbol{\rho}$ (on average, see below), which results in the input-state modification
\begin{equation}
    \boldsymbol{\rho} \to (\mathcal{L}_\tau \otimes \mathcal{L}_{\tau}) (\boldsymbol{\rho}).
    \label{eq:lossy and noisy states}
\end{equation}
This quantum channel can be represented via its Stinespring dilation as the interference of the state of interest with a vacuum state $\ket{0}\bra{0}_E$ on a beam splitter with transmissivity $\tau \in [0,1]$, after which the environmental mode $E$ is traced out, to wit,
\begin{equation}
    \mathcal{L}_\tau (\boldsymbol{\rho}_A) = \Tr_{E} \left\{ B_{AE}(\tau) \left( \boldsymbol{\rho}_A \otimes \ket{0} \bra{0}_E \right) B_{AE}^{\dagger} (\tau) \right\},
    \label{eq:LossyChannel}
\end{equation}
for subsystem $A$, and analogously for $B$.

\subsubsection{Noisy copies}
\label{subsubsec:Imperfect copies}
In real-world experiments, input-state and circuit parameters can vary from one experimental realization to another. The resulting variations among the state copies and subsequent experimental runs are commonly referred to as noise. We account for both effects by drawing any continuous parameter $x$, such as the complex phase $\alpha$ of the input state or the loss parameter $\tau$, from a narrow Gaussian distribution with standard deviation $\delta x \le x/10$ centered around the corresponding expected value $x$. 

We note that noise is sometimes accounted for by replacing the environmental vacuum state in~\eqref{eq:LossyChannel} with a thermal state, where the mean particle number acts as the noise parameter, corresponding to a noisy and lossy channel~\cite{Weedbrook2012}. Although we expect the resulting effects to be very similar, this modeling presents two practical drawbacks. First, mixing the input with a thermal state leads to (exponentially suppressed) excitations of arbitrarily high Fock states. Therefore, any reasonable numerical simulation of our method would require truncating the Fock spectrum above a specific threshold. Second, while this approach handles the noise during state preparation and within the optical circuit, it does not take into account fluctuations between different experimental realizations.

\subsubsection{Finite statistics}
\label{subsubsec:FiniteSampleEffects}
In every experiment, an observable of interest is ultimately estimated from a finite number of repetitions of the experiment, which can greatly influence the statistical significance of the expected outcome. In the context of entanglement detection, where the standard smoking-gun signal is the negativity of some witness function, finite statistics can significantly deteriorate detection capabilities by resulting in small confidence intervals.

We analyze the statistical behavior of the PT moments and the third-order criteria in terms of the means and variances of their estimators when sampled from $k$ independent and identically distributed (i.i.d.) experimental runs. We define the estimate of $p_n$ based on $k$ samples as
\begin{equation}
    p_n^{(k)} \equiv \frac{1}{k} \sum_{i=1}^k p_{n,(i)},
\end{equation}
where $p_{n,(i)} = \omega_n^{\sum_{j=1}^n(j-1)[N_{j,(i)}^A-N_{j,(i)}^B]}$, with $N_{j,(i)}^A$ representing the number of particles in Alice's $j$th mode during experiment $i$ (analogously for Bob) and $p_{n,(i)}$ being an i.i.d. complex random variable.

By linearity of the expectation value, we have 
\begin{equation}
    \text{E}[p_n^{(k)}] = \text{E}[p_{n,(i)}] = p_n.
\end{equation}
Further, $\text{Var}[p_{n,(i)}] = \text{E}[\abs{p_{n,(i)}}^2]-\abs{\text{E}[p_{n,(i)}]}^2 = 1-p_n^2$, where $\abs{p_{n,(i)}} = 1$ since $p_{n,(i)}$ is one of the $n$ roots of unity. Then, the variance of $p_n^{(k)}$ is simply
\begin{equation}
    \text{Var}[p_n^{(k)}] = \frac{1-p_n^2}{k}.
\end{equation}
Upon defining the estimators for the linear and quadratic criteria given in \eqref{eq:p3PPTCriteria} as
\begin{equation}
    \begin{split}
        W_l^{(k)} &\equiv p_3^{(k)} - \frac{3p_2^{(k)}-1}{2}, \\
        W_q^{(k)} &\equiv p_3^{(k)} - \frac{k |p_2^{(k)}|^2-1}{k-1},
    \end{split}
\end{equation}
which ensures unbiased estimators $\text{E}[W_l^{(k)}] = p_3-(3p_2-1)/2$ and $\text{E}[W_q^{(k)}] = p_3-p_2^2$, we obtain the variances
\begin{equation}
    \begin{split}
        \text{Var}[W_l^{(k)}] &= \frac{1-p_3^2}{k} + \frac{9}{4} \frac{1-p_2^2}{k}, \\
        \text{Var}[W_q^{(k)}] &= \frac{1-p_3^2}{k} + \frac{2(1-p_2^2)[1+(2k-3)p_2^2]}{k(k-1)}.
    \end{split}
    \label{eq:p3PPTCriteriaVariances}
\end{equation}

\subsection{NOON states -- a case study}
\label{subsec:Simulation}

\subsubsection{Analysis of individual errors}
We analyze the influence of all considered experimental imperfections on the prospects of witnessing higher-order NOON-state entanglement with our proposed method. To this end, we first incorporate losses following~\eqref{eq:LossyChannel}, which renders the underlying state mixed, thereby impeding entanglement detection. Expressions for the resulting state and its PT moments can be found in Appendix~\ref{app:lossy NOON states}. 

\begin{figure*}
    \centering
    \includegraphics[width=0.99\textwidth]{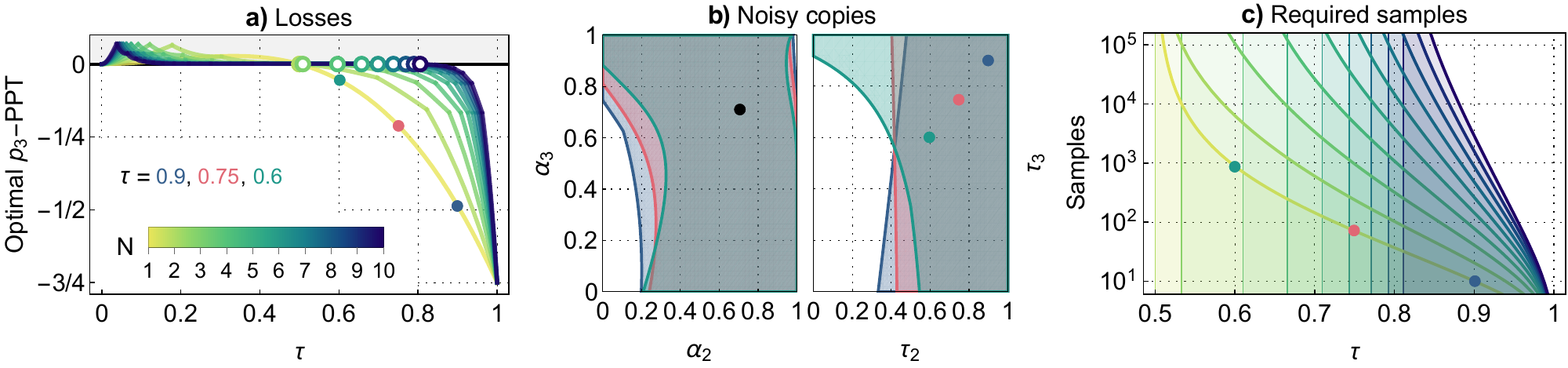}
    \caption{The three types of imperfections discussed in Sec.~\ref{subsec:Imperfections} applied to balanced ($\alpha = 1/\sqrt{2}$) NOON states. \textbf{a)} Optimal $p_3$-PPT criterion as a function of the loss parameter $\tau$ of the lossy channel $\mathcal{L}_\tau$. From yellow to blue, the NOON state's mode population number $N$ ranges from 1 to 10. Colored rings are placed at points where the criterion first flags entanglement, with a tolerance of up to $20\%$ losses until $N=10$ (note that $\tau > 1/2$ is required for all $N$). The colored dots at $N=1$ and $\tau=0.9$ (blue), $\tau=0.75$ (red) and $\tau = 0.6$ (green) allow for an intercomparison of typical loss values between the subfigures (other imperfections) and Fig.~\hyperref[fig:NOONStatesSimulation]{4} (full simulation). \textbf{b)} Regions of detected entanglement by the optimal $p_3$-PPT criterion for noisy copies of the first NOON state and the three typical loss parameters $\tau=0.9,0.75,0.6$. The first copy serves as a reference with the standard values for $\alpha_1 \equiv \alpha$ and $\tau_1 \equiv \tau$; the second copy is described by $\alpha_2$ and $\tau_2$ (for both $p_2$ and $p_3$), while $\alpha_3$ and $\tau_3$ parametrize the third copy used solely in the circuit for $p_3$. We illustrate the dependence on $\alpha$ in the left panel (with the black dot marking all $\alpha$'s being $1/\sqrt{2}$). Entanglement is also detected for large discrepancies, except when $\alpha_2$ approaches zero, i.e., when the second copy becomes separable. The right panel depicts the dependence on $\tau$, where entanglement is witnessed whenever $\tau_2 \gtrsim 1/2$. \textbf{c)} Number of samples required to detect entanglement via the quadratic criterion within one standard deviation for given loss $\tau$ and various mode occupations $N$ (vertical lines indicate minimal $\tau$ for infinite statistics). Even for $15 \%$ losses, entanglement is still witnessed up to $N=10$ with $10^5$ experimental runs only.}
    \label{fig:NOONStatesImperfections}
\end{figure*}

We show the absolute violation of the optimal $p_3$-PPT criterion as a function of the loss parameter $\tau$ for balanced ($\alpha = 1/\sqrt{2}$) NOON states up to $N=10$ in Fig.~\hyperref[fig:NOONStatesImperfections]{3a)}. As $N$ increases, fewer losses are tolerable for the state's entanglement to remain certified (see colored rings for the zero crossings). However, even for high occupations ($N=10$), our method detects entanglement up to $\sim 20\%$ loss (note that entanglement is never certified when losses exceed $50\%$ -- regardless of $N$ -- although the states remain entangled). 

To investigate noise effects (and to be able to sample the PT moments), we have to calculate the photon-number distributions at the outputs of the two circuits required for measuring $p_2$ and $p_3$, see Figs.~\hyperref[fig:Implementation]{1b)} and~\hyperref[fig:Implementation]{1c)}, respectively -- with generally different state and loss parameters for both subsystems and all copies. Note that this task resembles boson sampling~\cite{Aaronson2013}, which becomes inefficient on a classical computer rapidly with increasing $N$. Therefore, we restrict to $N=1$ for the analysis of noise and also for a full simulation of the experiment. We provide explicit expressions for the case where all parameters are equal and show the resulting output distributions for various loss parameters $\tau$ in Appendix~\ref{app:lossy NOON states output}.

We depict the impact of imperfect state copies on entanglement detection for $N=1$ in Fig.~\hyperref[fig:NOONStatesImperfections]{3b)}. In both panels, the first copy remains unaltered ($\alpha_1 = 1/\sqrt{2}$ and $\tau_1=0.9,0.75,0.6$ corresponding to the blue, red, and green regions, respectively), while the second and third copies are described by arbitrary $(\alpha_2, \tau_2)$ and $(\alpha_3, \tau_3)$, respectively. We examine noisy phases in the left panel and noisy losses in the right panel. In all cases, entanglement is certified substantially beyond the perfect-copy scenario (see points). We observe a maximum of $\approx 30\%$ allowed variations for $\tau = 0.6$ (green). This marks the minimum tolerance among the considered cases, thereby demonstrating strong robustness against noisy state copies.

At last, we study statistical effects. Using~\eqref{eq:p3PPTCriteriaVariances}, we compute the least amount of samples needed to witness entanglement with the quadratic criterion within one standard deviation [see Fig.~\hyperref[fig:NOONStatesImperfections]{3c)}]. For the Bell state (${N=1}$), only $\approx 10$ samples are required at $10 \%$ losses (blue point), while $\approx 10^3$ samples are still sufficient in the very lossy regime $\tau = 0.6$ (green point). Remarkably, our method certifies higher-order NOON entanglement up to $N=10$ at $\approx 15\%$ losses for a reasonable measurement budget of $10^5$ experimental runs, showcasing its efficiency also in the strongly non-Gaussian regime.

\subsubsection{Full simulation}
To complete the picture, we combine all sources of imperfections by sampling from the photon-number-output distribution of the lossy and noisy Bell state (${N=1, \alpha = 1/\sqrt{2}}$). For each individual copy and each experimental run, we draw the phases $\alpha$ and the losses $\tau$ from Gaussian distributions centered around $\alpha = 1/\sqrt{2}$ and $\tau=0.9,0.75,0.6$ (blue, red, green), respectively, and with standard deviations $5 \%$ and $0.05$, respectively, to represent the overall noisiness of a real-world experiment. Then, we compute the optimal $p_3$-PPT criterion for a given number of samples, which we repeat $500$ times to gather statistics on the mean and its error.

\begin{figure}[t!]
    \centering
    \includegraphics[width=0.85\columnwidth]{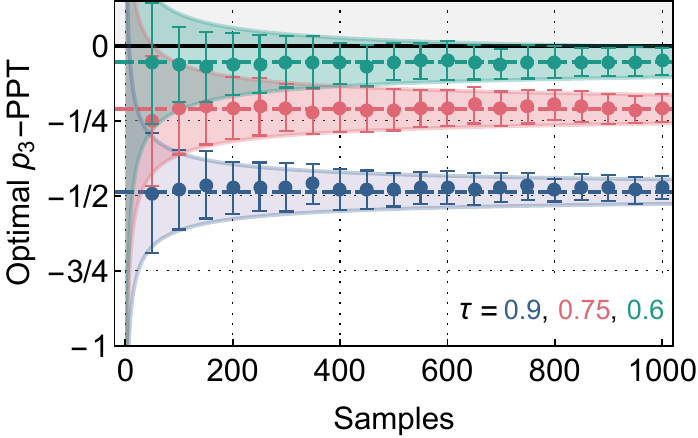}
    \caption{Simulated experiment for detecting the entanglement of the first NOON state ($N=1$), including the effects of losses, noisy copies, and finite statistics. We plot the absolute violation of the optimal $p_3$-PPT criterion over the number of samples. The data points and their error bars representing one standard deviation are obtained after averaging over 500 repetitions. The dashed horizontal lines depict the perfect-copy (but still lossy) case. At the same time, the surrounding shaded areas indicate the one standard deviation intervals from the analytical error model for the linear criterion [see~\eqref{eq:p3PPTCriteriaVariances}]. Given that all curves fall below zero for less than $10^3$ samples, our method is capable of detecting strongly non-Gaussian entanglement in a close-to-experimental setting.}
    \label{fig:NOONStatesSimulation}
\end{figure}

We showcase the performance of the optimal criterion as a function of the sample size in Fig.~\ref{fig:NOONStatesSimulation}. The simulated data points fluctuate around the analytical predictions (dashed lines, with losses taken into account), and their one-standard-deviation intervals fall within the analytical error bounds of the linear witness [cf.~\eqref{eq:p3PPTCriteriaVariances}]. For a detailed comparison of the three third-order criteria and the error models without additional noise effects, we refer to Appendix~\ref{app:NOONErrorModels}. We observe that entanglement is witnessed within one standard deviation up to $40 \%$ loss using $10^3$ samples only (green points and curves), which is well within the capabilities of current experimental technologies. To conclude, our method remains robust against the typical errors occurring in a realistic experimental setting for the studied cases. 

\section{Discussion}
\label{sec:Discussion}

In summary, we generalized entanglement criteria based on moments of the density operator's partial transpose to continuous variables and exploited the multicopy approach in order to access low-order PT moments. Our measurement schemes
use linear interferometers and a few replicas of the state, together with particle-number-resolving detectors. Compared to standard approaches, our method exhibits strong advantages regarding both detection capability -- especially in the strongly non-Gaussian regime -- and robustness against experimental imperfections, including losses, noise, and finite statistics. In particular, we demonstrated that our criteria can witness higher-order NOON-type entanglement under realistic experimental conditions, highlighting their usefulness in practical applications such as photonic quantum processors.  

In comparison, in~\cite{Gessner2016}, a metrological witness is proposed. It relies on local variances and provides a necessary and sufficient criterion for pure-state entanglement, but does not address the role of imperfections or the regime of strongly non-Gaussian entanglement. More recently, in~\cite{Gao2024}, neural networks are employed to certify non-Gaussian-state entanglement using homodyning. They also account for losses, using the same model as in this work, in the context of single-photon-subtracted and entangled coherent states, whereas our focus is on NOON-state entanglement. A more detailed comparative study of these different approaches would be an interesting direction for future work.

The need for interferometric stability over several state copies -- though not necessarily perfect -- is arguably the most significant experimental burden of our method. Hence, we believe that it is of particular interest to extend our criteria, and more generally the multicopy method, to arbitrary ancilla states, which would further reduce this experimental hurdle. We refer to~\cite{Callus2025} for a recently introduced entanglement detection scheme where higher-order correlations of a given bipartite state are accessed via its interference with an arbitrary two-mode ancilla.

A remaining challenge lies in generating non-Gaussian states. While probabilistic preparation makes producing several copies more demanding, not all schemes are probabilistic~\cite{Pasharavesh2024}, and the success rate depends on the specific state~\cite{Su2019}. Moreover, probabilistic preparation is effectively equivalent to reduced statistics, and our results show that only about $10^3$ samples are sufficient to witness NOON-state entanglement within realistic experimental conditions. 

We consider several other directions to be interesting for future work. First, one may extend our method to capture moments of other operators resulting from applying positive but not completely positive maps. This includes, for instance, the realignment map~\cite{Rudolph2005,PhysRevLett.111.190501,PhysRevA.104.022427}, for which moment-based criteria have been derived recently in~\cite{Aggarwal2024}. Compared to the PT-moment approach, this could enable the detection of bound entanglement, which is known to be scarce in CV systems~\cite{Horodecki2003}. 

Second, since we have exclusively discussed a two-mode bipartition, one may generalize our setup to both multimode and multipartite cases to analyze many-body and/or multipartite entanglement. The extension to a multimode bipartition is straightforward: The optical circuits for measuring PT moments have to be applied to each mode individually, upon which the particle numbers within one subsystem are summed up. As the number of copies remains constant, the total system size still scales linearly with the number of target state modes. Further, a possible generalization that would allow one to certify multipartite entanglement could be based on the corresponding variant of the PPT criterion~\cite{Jungnitsch2011}.

Third, in a somewhat different direction, one may ask whether the preferred readout scheme in the discrete-variable case, that is, randomized measurements in spin bases, can be transferred directly to the CV regime via the Jordan-Schwinger map~\cite{Haas2024e} (see also~\cite{Gandhari2024,Becker2024} for CV shadow tomography). This would not only constitute another way of accessing PT moments, but could be a general alternative to the multicopy method. Since the required number of randomized measurements diverges for an infinite-dimensional Hilbert space, one would have to impose a finite energy (or particle-number) cutoff, which is fairly reasonable from an experimental perspective and a standard procedure in CV quantum information~\cite{Weedbrook2012}. 

\section*{Acknowledgements}
We thank Elena Callus and Martin Gärttner for useful discussions and comments on the manuscript, as well as Célia Griffet for preliminary work that led to this project. S.D. is a FRIA grantee of the F.R.S.--FNRS. N.J.C. and T.H. acknowledge support from the F.R.S.--FNRS under project CHEQS within the Excellence of Science (EOS) program. N.J.C. and T.H. were further supported by the European Union under project ShoQC within the ERA-NET Cofund in Quantum Technologies (QuantERA) program.

\section*{Data availability and source code}
We provide the code used for generating Figs. 2--4 and 6--8 via the GitHub repository~\cite{GitHub}.


\appendix

\section{Gaussian states}

\subsection{Two-mode squeezed state at finite temperature}
\label{app:TMSState}
To illustrate the breakdown of the various PT-moment-based criteria for sufficiently mixed Gaussian states, we consider a two-mode squeezed state at finite temperature. Starting from a product of two identical thermal states $\boldsymbol{\rho}_{\bar{n}} = \sum_{n=0}^\infty \frac{{\bar{n}}^n}{({\bar{n}}+1)^{(n+1)}} \ket{n}\bra{n}$ with mean particle number $\bar{n} \ge 0$, this state is generated via the two-mode squeezer unitary $\boldsymbol{U}_{r} = \text{exp}[r(\boldsymbol{a}\boldsymbol{b}-\boldsymbol{a}^\dagger\boldsymbol{b}^\dagger)/2]$ with squeezing parameter $r \in [0, \infty)$, resulting in 
\begin{equation}
    \boldsymbol{\rho}_{\bar{n},r} = \boldsymbol{U}_{r}(\boldsymbol{\rho}_{\bar{n}} \otimes \boldsymbol{\rho}_{\bar{n}})\boldsymbol{U}_{r}^\dagger.
    \label{eq:TMSFiniteTemperature}
\end{equation}
One can easily find that the symplectic eigenvalues of the partially transposed state are $\tilde\nu_1 = e^{-2r}\left(2{\bar{n}}+1\right)$ and $ \tilde\nu_2 = e^{2r}\left(2{\bar{n}}+1\right)$, such that the state's purity depends on the mean particle number only, i.e., $p_2 = 1/\left(2{\bar{n}}+1\right)^2$.

When $\bar{n}=0$,~\eqref{eq:TMSFiniteTemperature} reduces to the pure two-mode squeezed vacuum state [see dashed black curves in Figs.~\hyperref[fig:Benchmarks]{2b)} and~\hyperref[fig:Benchmarks]{2c)}], which is entangled for any finite squeezing $r>0$. Upon increasing $\bar{n}$, the state becomes more mixed, with purity $p_2=1/2$ (see dashed yellow curves) being attained for $\bar{n}=(\sqrt{2}-1)/2 \approx 0.207$ particles per mode. In this regime, sufficient squeezing $r > \ln (2\bar{n}+1)/2$ is required for the state to be entangled at all. For the above case, this entails $r > \ln (2)/4 \approx 0.173$ or $p_3 < 4/13$ (see yellow squares). However, all third-order criteria require $r \gtrsim 0.363$ or $p_3 < 1/4$ to detect entanglement (see yellow diamonds). Note that physicality translates into $p_3 \le 16/49$ for $p_2 = 1/2$ [see yellow triangle in Fig.~\hyperref[fig:Benchmarks]{2b)}], showing that the range of values for $p_3$ compatible with a separable state is small compared to an entangled state [see the light gray region in the inset of Fig.~\hyperref[fig:Benchmarks]{2b)}].

\subsection{Higher-order criteria}
\label{app:GaussianStates}
We analyze higher-order $p_n$-PPT criteria from the Stieltjes approach [cf.~\eqref{eq:pnPPTCriteria}] for Gaussian states in Fig.~\ref{fig:GaussianHigherMoments}. For the two-mode squeezed state at finite temperature considered in Appendix~\ref{app:TMSState}, the $p_n$-PPT criteria for $n=5$ (green curve) and $n=7$ (red curve) flag entanglement already when $r \gtrsim 0.221$ (yellow squares) and $r \gtrsim 0.187$ (yellow triangles), respectively, compared to $r \gtrsim 0.363$ (yellow diamonds) for $n=3$.

\begin{figure}[t!]
    \centering
    \includegraphics[width=0.55\columnwidth]{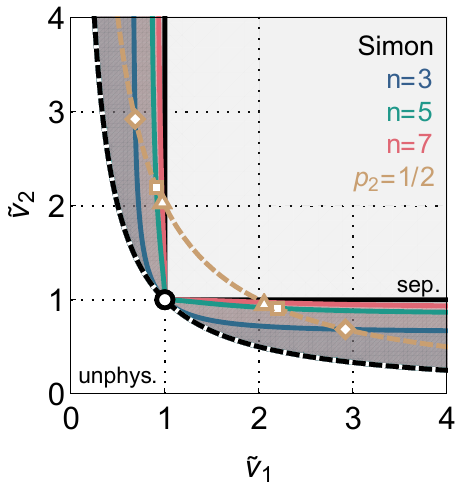}
    \caption{Gaussian entanglement detected by the $p_n$-PPT criteria \eqref{eq:pnPPTCriteria} of order $n=3$ (blue), $n=5$ (petrol), and $n=7$ (red), analogous to Fig.~\hyperref[fig:Benchmarks]{2a)}. More mixed entangled states are detected for increasing $n$, indicating convergence toward Simon's criterion in the limit $n \to \infty$.}
    \label{fig:GaussianHigherMoments}
\end{figure}

\section{Mixed entangled cat states}
\label{app:CatStates}
To evaluate the PT moments for the class of states~\eqref{eq:CatStates}, we make abundantly use of the coherent-state inner product $\braket{\alpha|\beta} = e^{-\frac{1}{2} (|\alpha-\beta|^2 + \alpha \beta^* - \alpha^* \beta)}$, which leads to
\begin{equation}
    \begin{split}
        p_2 &= \frac{2}{C^2} \, \Big\{[1+(1-z)^2][1+e^{-4(|\alpha|^2+|\beta|^2)}] \\ &\hspace{1.2cm}\pm4(1-z)e^{-2(|\alpha|^2+|\beta|^2)} \Big\}
    \end{split}
\end{equation}
and
\begin{equation}
    \begin{split}
        \hspace{-0.06cm}p_3 &= \frac{2}{C^3} \, \Big\{1+3e^{-4(|\alpha|^2+|\beta|^2)}  \\ 
        &\hspace{0.4cm} \pm3(1-z) e^{-2(|\alpha|^2+|\beta|^2)} \big[ 2 + e^{-4|\alpha|^2} + e^{-4 |\beta|^2} \big] \\ 
        &\hspace{0.4cm}+3(1-z)^2 e^{-4(|\alpha|^2+|\beta|^2)} \big[ 2+e^{4|\alpha|^2} + e^{4|\beta|^2} \big] \\
        &\hspace{0.4cm}\pm(1-z)^3 e^{-6(|\alpha|^2+|\beta|^2)} \big[ 1 + 3e^{4(|\alpha|^2+|\beta|^2)}\big] \Big\}.
    \end{split}
\end{equation}

\section{High Harmonic Generation (HHG)}
\label{app:HHGStates}
The reduced state of the driving field and one harmonic mode for equal harmonic shifts ${\chi \equiv \chi_q = (2/\sqrt{N^2 + 2N - 3}) \, \delta \alpha}$ reads
\begin{equation}
    \begin{split}
        \boldsymbol{\rho} &= \frac{1}{C} \bigg[ \ket{\alpha-\delta\alpha,\chi}\bra{\alpha-\delta\alpha,\chi} \\
        &\hspace{1cm}- e^{-\frac{N^2+6N-11}{2(N^2+2N-3)}|\delta\alpha|^2} \ket{\alpha-\delta\alpha,\chi}\bra{\alpha,0} \\
        &\hspace{1cm}- e^{-\frac{N^2+6N-11}{2(N^2+2N-3)}|\delta\alpha|^2} \ket{\alpha,0}\bra{\alpha-\delta\alpha,\chi} \\
        &\hspace{1cm}+ e^{-\frac{N^2+4N-5}{N^2+2N-3}|\delta\alpha|^2} \ket{\alpha,0}\bra{\alpha,0} \bigg],
    \end{split}
\end{equation}
where we assumed that $\alpha, \delta \alpha$, and $\chi$ are real without loss of generality. The purity evaluates to
\begin{equation}
    \begin{split}
        p_2 &=\frac{1}{C^2}\bigg[1-2\left( 1-C \right)\left(2 - e^{-(|\delta\alpha|^2+|\chi|^2)} \right) \\
        &\hspace{1.2cm}+ \left( 1-C \right)^2 \left(2 e^{|\delta\alpha|^2+|\chi|^2}-1\right)\bigg],
    \end{split}
\end{equation}
while the third PT moment is
\begin{equation}
    \begin{split}
        p_3 &= \frac{1}{C^3} \bigg[ 1-3(1-C)\left(2-e^{-(|\delta\alpha|^2+|\chi|^2)} \right) \\
        &\hspace{1.2cm}+ 3(1-C)^2\Big( 2 +e^{-(|\delta\alpha|^2+|\chi|^2)} \\
    &\hspace{3.3cm}+e^{|\delta\alpha|^2} -2e^{-|\delta\alpha|^2} \\
        &\hspace{3.3cm}+e^{|\chi|^2}-2e^{-|\chi|^2}\Big) \\
        &\hspace{1.2cm}-(1-C)^3\Big(1 + 6e^{|\delta\alpha|^2+|\chi|^2}\\
        &\hspace{3.1cm}-3e^{|\delta\alpha|^2}-3e^{|\chi|^2}\Big) \bigg].
    \end{split}
\end{equation}

\section{NOON states}
\label{app:NOONStates}

\subsection{Mode-operator moments}
\label{app:NOONMoments}
We compute the set of constraint mode-operator moments for the class of NOON states. To this end, we normal order~\eqref{eq:NOONState}, for which we use the identity $\ket{0}\bra{0}= :e^{-\boldsymbol{a}^{\dagger} \boldsymbol{a}}: = \sum_j \boldsymbol{a}^{\dagger j} \boldsymbol{a}^j (-1)^j/(j!)$. We end up with
\begin{equation}
    \begin{split}
        \hspace{-0.17cm}\rho_{i j k l} &= \abs{\alpha}^2 \frac{(-1)^{i+k-N}}{N! \, (i-N)! \, k!} \delta_{i j} \delta_{k l} \Theta (i-N) \\
        &\hspace{0.3cm} +\abs{\beta}^2 \frac{(-1)^{i+k-N}}{N! \, i! (k-N)!} \delta_{i j} \delta_{k l} \Theta (k-N) \\
        &\hspace{0.3cm} + \alpha \beta^* \frac{(-1)^{j+k}}{N! \, j! \, k!} \delta_{i \, j+N} \, \delta_{k \, l-N} \Theta (i-N) \Theta (l-N) \\
        &\hspace{0.3cm} + \alpha^* \beta \frac{(-1)^{i+l}}{N! \, i! \, l!} \delta_{i \, j-N} \, \delta_{k \, l+N} \Theta (j-N) \Theta (k-N),
    \end{split}
\end{equation}
where $\Theta (x)$ denotes the Heaviside $\Theta$ function (with the convention $\Theta (0)=1$ understood). This leads to the expansion for the purity
\begin{equation}
    \begin{split}
        p_2 &\sim \frac{\abs{\alpha}^2}{N!} \left[ \braket{\boldsymbol{a}^{\dagger N} \boldsymbol{a}^{N}} + \braket{\boldsymbol{a}^{\dagger (N + \cdots)} \boldsymbol{a}^{N + \cdots} \boldsymbol{b}^{\dagger (0+\cdots)} \boldsymbol{b}^{0 + \cdots}} \right] \\
        &\hspace{0.3cm}+ \frac{\abs{\beta}^2}{N!} \left[\braket{\boldsymbol{b}^{\dagger N} \boldsymbol{b}^{N}} + \braket{\boldsymbol{a}^{\dagger (0 + \cdots)} \boldsymbol{a}^{0 + \cdots} \boldsymbol{b}^{\dagger (N+\cdots)} \boldsymbol{b}^{N + \cdots}} \right] \\
        &\hspace{0.3cm}+ \frac{\alpha \beta^*}{N!} \left[ \braket{\boldsymbol{a}^{\dagger N} \boldsymbol{b}^{N}} + \braket{\boldsymbol{a}^{\dagger (N+\cdots)} \boldsymbol{a}^{0+\cdots} \boldsymbol{b}^{\dagger (0+\cdots)} \boldsymbol{b}^{N+\cdots}} \right]\\
        &\hspace{0.3cm}+ \frac{\alpha^* \beta}{N!} \left[ \braket{\boldsymbol{a}^{N} \boldsymbol{b}^{\dagger N}} + \braket{\boldsymbol{a}^{\dagger (0+\cdots)} \boldsymbol{a}^{N+\cdots} \boldsymbol{b}^{\dagger (N+\cdots)} \boldsymbol{b}^{0+\cdots}} \right],
    \end{split}
\end{equation}
where center dots indicates higher-order contributions. Only the first term in each row survives, and the purity reduces to
\begin{equation}
    \begin{split}
        p_2 &= \frac{1}{N!} \Big[ \abs{\alpha}^2 \braket{\boldsymbol{a}^{\dagger N} \boldsymbol{a}^{N}} + \abs{\beta}^2  \braket{\boldsymbol{b}^{\dagger N} \boldsymbol{b}^{N}} \\
        &\hspace{0.8cm}+ \alpha \beta^* \braket{\boldsymbol{a}^{\dagger N} \boldsymbol{b}^{N}} + \alpha^* \beta  \braket{\boldsymbol{a}^{N} \boldsymbol{b}^{\dagger N}}\Big].
    \end{split}
    \label{eq:p2NOONState2}
\end{equation}
Similarly, we find for the third PT moment
\begin{equation}
    p_3 = \frac{1}{N!} \left[ \abs{\alpha}^4 \braket{\boldsymbol{a}^{\dagger N} \boldsymbol{a}^{N}} + \abs{\beta}^4  \braket{\boldsymbol{b}^{\dagger N} \boldsymbol{b}^{N}} \right].
    \label{eq:p3NOONState2}
\end{equation}
It is straightforward to verify consistency of the latter two equations with $p_2=1$ and~\eqref{eq:p3NOONState} since $\braket{\boldsymbol{a}^{\dagger N} \boldsymbol{a}^{N}}= \abs{\alpha}^2 N!$ (analogously for $\boldsymbol{b}$), and $\braket{\boldsymbol{a}^{N} \boldsymbol{b}^{\dagger N}}=\braket{\boldsymbol{a}^{\dagger N} \boldsymbol{b}^{N}}^*= \alpha \beta^*N!$ for NOON states. 

Equations~\eqref{eq:p2NOONState2} and~\eqref{eq:p3NOONState2} nicely demonstrate the main strength of PT-moment-based entanglement criteria: In contrast to entropic and Shchukin-Vogel-type criteria, where higher-order mode-operator moments are exponentially suppressed or hard to access, respectively, PT moments often select the right (and in principle arbitrarily high) mode-operator moments that contain the necessary information about quantum correlations. 

\subsection{Lossy NOON states}
\label{app:lossy NOON states}
Since we account for noisy copies by fluctuating state parameters, we consider in general different losses for Alice and Bob. Then, the lossy NOON state reads
\begin{equation}
    \boldsymbol{\rho} =  (\mathcal{L}_{\tau_A} \otimes \mathcal{L}_{\tau_B})\,(\ket{\psi}\bra{\psi}),
\end{equation}
where $\ket{\psi}$ is the lossless NOON state as defined in~\eqref{eq:NOONState}. Direct calculation leads to
\begin{equation}
    \begin{split}
        \boldsymbol{\rho} =& |\alpha|^2 \sum_{k=0}^N \binom{N}{k} \tau_A^{(N-k)}(1-\tau_A)^k\ket{N-k,0}\bra{N-k,0}\\ &+ \sqrt{\tau_A\tau_B}^N \big(\alpha \beta^* \ket{N,0}\bra{0,N} +\alpha^* \beta \ket{0,N}\bra{N,0}\big)\\ &+ |\beta|^2 \sum_{k=0}^N \binom{N}{k} \tau_B^{(N-k)}(1-\tau_B)^k\ket{0,N-k}\bra{0,N-k}.
    \end{split}
\end{equation}
The PT moments evaluate to
\begin{equation}
    \begin{split}
        p_2 &= |\alpha|^4 \sum_{k=0}^N \binom{N}{k}^2 \tau_A^{2(N-k)}(1-\tau_A)^{2k}\\ &+ 2|\alpha|^2|\beta|^2\big(\tau_A^N \tau_B^N+(1-\tau_A)^N(1-\tau_B)^N\big)\\ &+ |\beta|^4 \sum_{k=0}^N \binom{N}{k}^2 \tau_B^{2(N-k)}(1-\tau_B)^{2k}
    \end{split}
\end{equation}
and
\begin{equation}
    \begin{split}
        p_3 &= |\alpha|^6 \sum_{k=0}^N \binom{N}{k}^3 \tau_A^{3(N-k)}(1-\tau_A)^{3k}\\ &+ 3|\alpha|^2|\beta|^2\bigg(|\alpha|^2\tau_A^N\tau_B^N(1-\tau_A)^N  + |\beta|^2\tau_A^N\tau_B^N(1-\tau_B)^N  \\ &+|\alpha|^2(1-\tau_A)^{2N}(1-\tau_B)^{N}+|\beta|^2(1-\tau_A)^{N}(1-\tau_B)^{2N}\bigg)\\ &+ |\beta|^6 \sum_{k=0}^N \binom{N}{k}^3 \tau_B^{3(N-k)}(1-\tau_B)^{3k},
    \end{split}
\end{equation}
where we considered the copies to be identical for brevity. Please note that the expressions underlying Sec.~\ref{subsec:Simulation} describe the more general case of unequal copies.

\subsection{Particle-number output distributions}
\label{app:lossy NOON states output}

\begin{figure*}[p]
    \centering
    \includegraphics[width=0.95\textwidth]{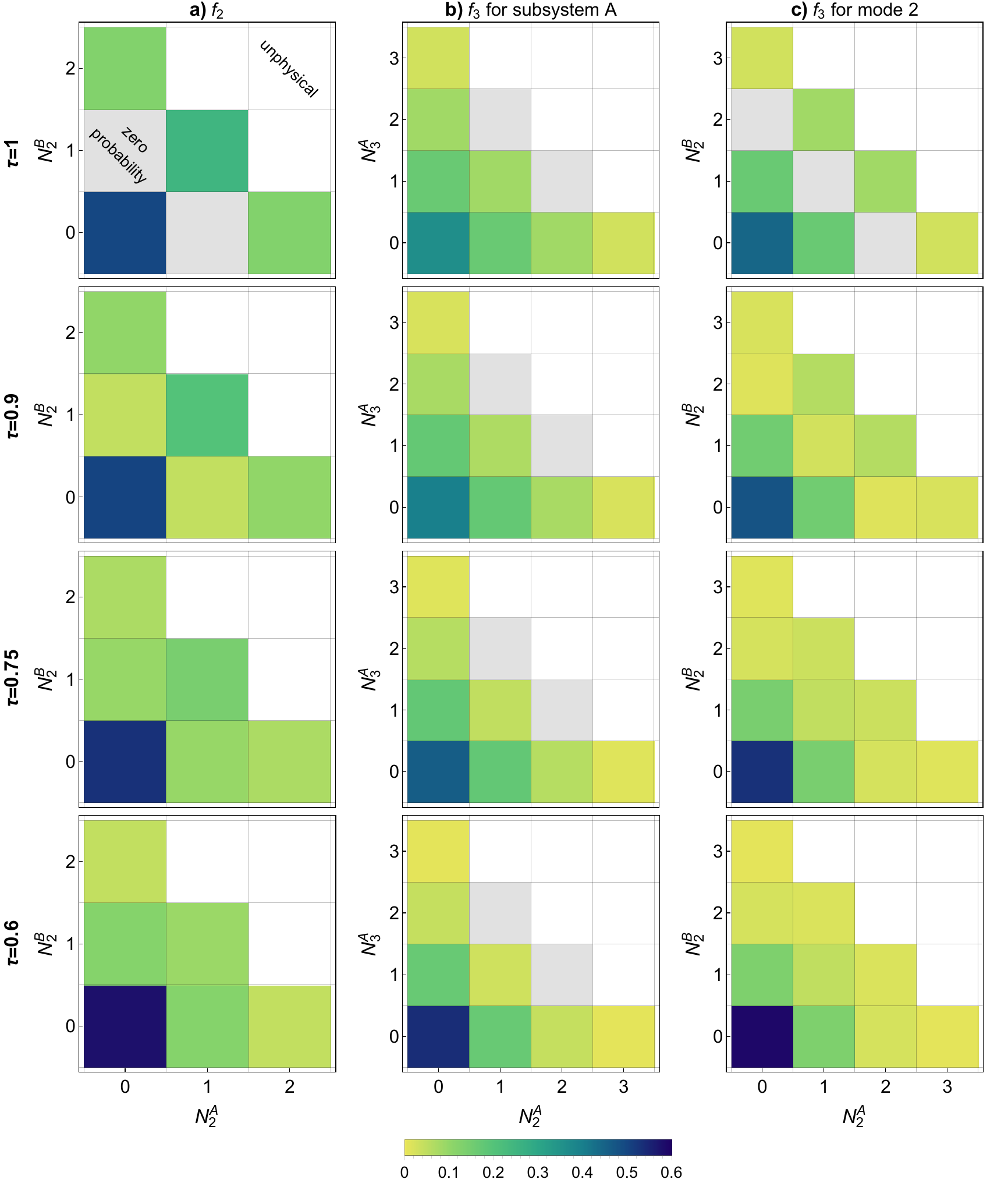}
    \caption{Particle-number output distributions for the balanced ($\alpha = 1/\sqrt{2}$) NOON state of order $N=1$ for various losses $\tau$. \textbf{a)} shows $f_2$ from the purity circuit (see Table~\ref{tab:PhotonNumberDistributionsp2Modified} for analytic expressions). \textbf{b)} and \textbf{c)} depict $f_3$ from the $p_3$ circuit for subsystem $A$ and in the second mode, respectively (see Table~\ref{tab:PhotonNumberDistributionsp3Modified}). When the total particle number at the output exceeds its input value, any corresponding configuration is unphysical (white elements). In some cases, physical configurations attain zero probability (gray elements), e.g., when the state is pure ($\tau = 1$) and the total particle number is odd (see the top left panel).}
    \label{fig:PhotonNumberDistributions}
\end{figure*}

We show the particle-number output distributions $f_2 (N^A_2,N_2^B)$ and $f_3 (N^A_2,N_2^B,N^A_3,N_3^B)$ of the Fourier interferometers of orders 2 [see Fig.~\hyperref[fig:Implementation]{1b)}] and 3 [see Fig.~\hyperref[fig:Implementation]{1c)}] in Tables~\ref{tab:PhotonNumberDistributionsp2Modified} and~\ref{tab:PhotonNumberDistributionsp3Modified}, respectively. Again, we present expressions for equal copies for brevity. Note that at most two (three) particles can be detected at the outputs for two (three) NOON input states. Note also that for $\tau = 1$, i.e., when the state is pure, contributions with odd total particle number $N^A_2 + N^B_2$ vanish since all individual parities must be even to average to $p_2=1$. Additionally, we provide graphical visualizations of the output distributions at the typical loss levels considered in the main text, i.e., $\tau=0.9,0.75,0.6$, together with the pure-state case $\tau=1$ [see Fig.~\ref{fig:PhotonNumberDistributions}].

\clearpage

\renewcommand{\arraystretch}{0.9}

\begin{table}[!t]
    \centering
    \setlength{\tabcolsep}{4pt} 
    \begin{tabular}{c c c c}
    \toprule
    $N_{\text{tot}}$ & $N_2^A$ & $N_2^B$ & $f_2$ \\
    \midrule
    \multirow{1}{*}{\bf{0}} 
      & 0 & 0 & $1 - \frac{\tau}{2} (2-\tau)$ \\
    \midrule
    \multirow{2.5}{*}{\bf{1}} 
      & 1 & 0 & $\tau (1 - \tau) \abs{\alpha}^2$ \\
      \cmidrule(lr){2-4}
      & 0 & 1 & $\tau (1 - \tau) \abs{\beta}^2$ \\
    \midrule
    \multirow{4.35}{*}{\bf{2}} 
      & 1 & 1 & $\tau^2 \abs{\alpha}^2 \abs{\beta}^2$ \\
      \cmidrule(lr){2-4}
      & 2 & 0 & $\frac{\tau^2}{2} \abs{\alpha}^4$ \\
      \cmidrule(lr){2-4}
      & 0 & 2 & $\frac{\tau^2}{2} \abs{\beta}^4$ \\
    \bottomrule
    \end{tabular}
    \caption{Particle-number output distribution $f_2$ of the circuit in Fig.~\hyperref[fig:Implementation]{1b)} for a lossy NOON state ($N=1$) over all possible particle numbers $N_2^A$, $N_2^B$ grouped by total particle number $N_{\text{tot}} = N_2^A + N_2^B$.}
    \label{tab:PhotonNumberDistributionsp2Modified}
\end{table}

\subsection{Statistical error models}
\label{app:NOONErrorModels}
\begin{figure*}[p]
    \centering

    \includegraphics[width=0.99\textwidth]{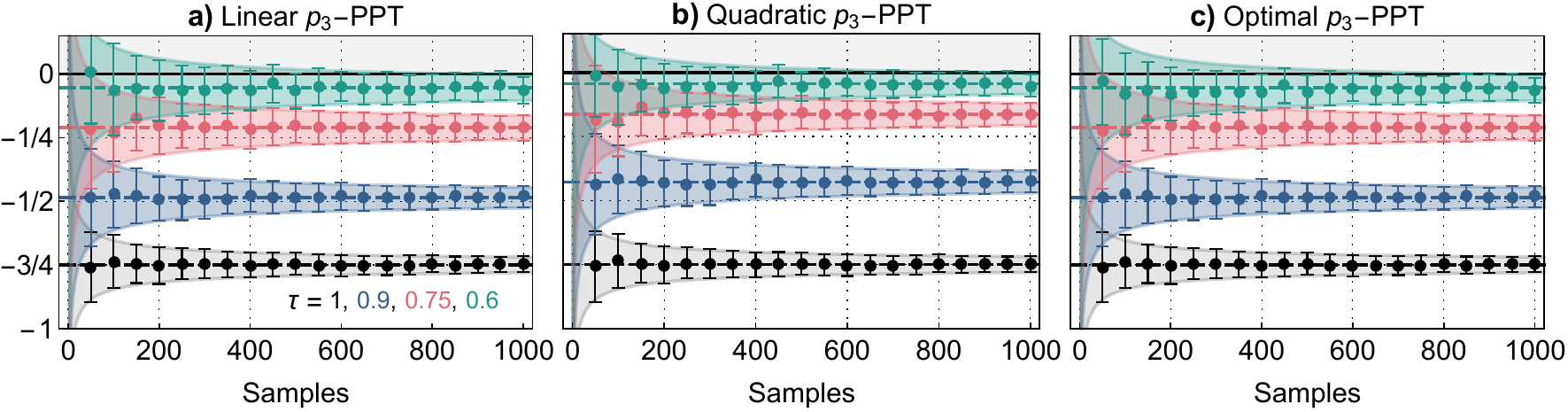}
    \caption{Numerical validation of the statistical error models for the three $p_3$-PPT criteria in the case of the noiseless and balanced ($\alpha = 1/\sqrt{2}$) NOON state with mode population $N=1$, considered for the three typical losses $\tau=0.9,0.75,0.6$ (blue, red, green) together with the lossless case $\tau=1$ (black). See Fig.~\ref{fig:NOONStatesSimulation} for reference.}
    \label{fig:NOONStatesErrorModel1}

    \vspace{3em}

    \includegraphics[width=0.40\textwidth]{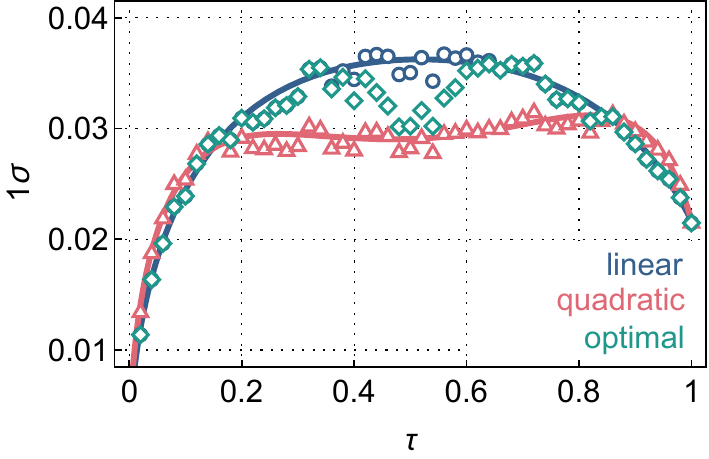}
    \caption{Complementary analysis to Fig.~\ref{fig:NOONStatesErrorModel1} in terms of the standard deviation as a function of the loss $\tau$. Analytic curves correspond to~\eqref{eq:p3PPTCriteriaVariances}. The criteria are sampled $2 \times 10^3$ times, which we repeat $500$ times to obtain their standard deviations (points). The optimal criterion's error mostly resembles the linear one's behavior as $p_2 \geq 1/2$ with equality if and only if $\tau = 1/2$.}
    \label{fig:NOONStatesErrorModel2}

\end{figure*}

We validate the consistency between the statistical error models discussed in Sec.~\ref{subsubsec:FiniteSampleEffects} and the numerical sampling of the third-order criteria for the noiseless and balanced ($\alpha = 1/\sqrt{2}$) NOON state of order $N=1$ in Fig.~\ref{fig:NOONStatesErrorModel1}. As a complementary check, we plot the standard deviations obtained in~\eqref{eq:p3PPTCriteriaVariances} relative to the loss $\tau$ in Fig.~\ref{fig:NOONStatesErrorModel2}. The numerical data aligns well with the theoretical predictions for the linear and quadratic criteria. The optimal criterion matches the linear one when $\tau \lesssim 0.4$ and $\tau \gtrsim 0.6$ (which corresponds to $p_2 > 1/2$), but approaches the quadratic one towards $\tau \approx 1/2$ (for which $p_2 \approx 1/2$). As the linear criterion's error is an upper bound for the optimal criterion's error, it is justified to work with the linear error model for the optimal criterion when $\tau \gtrsim 0.6$, which comprises all cases considered in the main text, in particular, Figs.~\hyperref[fig:NOONStatesImperfections]{3c)} and~\ref{fig:NOONStatesSimulation}.

\begin{table}[!t]
    \centering
    \setlength{\tabcolsep}{4pt}
    \begin{tabular}{c c c c c c}
    \toprule
    $N_{\text{tot}}$ & $N_2^A$ & $N_2^B$ & $N_3^A$ & $N_3^B$ & $f_3$ \\
    \midrule
    \multirow{1}{*}{\bf{0}} 
      & 0 & 0 & 0 & 0 & $1 - \frac{1}{9} \tau [18 + \tau (-15 + 4 \tau)]$ \\
    \midrule
    \multirow{4.5}{*}{\bf{1}}
      & 1 & 0 & 0 & 0 & \multirow{2}{*}{$\frac{1}{3} \abs\alpha^2 (1 - \tau) \tau (3 - 2 \tau)$} \\
      & 0 & 0 & 1 & 0 & \\
      \cmidrule(lr){2-6}
      & 0 & 1 & 0 & 0 & \multirow{2}{*}{$\frac{1}{3} \abs\beta^2 (1 - \tau) \tau (3 - 2 \tau)$} \\
      & 0 & 0 & 0 & 1 & \\
    \midrule
    \multirow{12.4}{*}{\bf{2}} 
      & 1 & 1 & 0 & 0 & \multirow{2}{*}{$\frac{4}{3} \abs\alpha^2 \abs\beta^2 (1 - \tau) \tau^2$} \\
      & 0 & 0 & 1 & 1 & \\
      \cmidrule(lr){2-6}
      & 1 & 0 & 0 & 1 & \multirow{2}{*}{$\frac{1}{3} \abs\alpha^2 \abs\beta^2 \tau^2$} \\
      & 0 & 1 & 1 & 0 & \\
      \cmidrule(lr){2-6}
      & 1 & 0 & 1 & 0 & $\frac{1}{3} \abs\alpha^4 \tau^2$ \\
      \cmidrule(lr){2-6}
      & 0 & 1 & 0 & 1 & $\frac{1}{3} \abs\beta^4 \tau^2$\\
      \cmidrule(lr){2-6}
      & 2 & 0 & 0 & 0 & \multirow{2}{*}{$\frac{2}{3} \abs\alpha^4 (1 - \tau) \tau^2$} \\
      & 0 & 0 & 2 & 0 & \\
      \cmidrule(lr){2-6}
      & 0 & 2 & 0 & 0 & \multirow{2}{*}{$\frac{2}{3} \abs\beta^4 (1 - \tau) \tau^2$} \\
      & 0 & 0 & 0 & 2 & \\
    \midrule
    \multirow{21.7}{*}{\bf{3}} 
      & 1 & 2 & 0 & 0 & \multirow{2}{*}{$\frac{2}{3} \abs\alpha^2 \abs\beta^4 \tau^3$} \\
      & 0 & 0 & 1 & 2 & \\
      \cmidrule(lr){2-6}
      & 2 & 1 & 0 & 0 & \multirow{2}{*}{$\frac{2}{3} \abs\alpha^4 \abs\beta^2 \tau^3$} \\
      & 0 & 0 & 2 & 1 & \\
      \cmidrule(lr){2-6}
      & 1 & 0 & 2 & 0 & \multirow{12}{*}{$0$} \\
      & 2 & 0 & 1 & 0 & \\
      & 0 & 1 & 0 & 2 & \\
      & 0 & 2 & 0 & 1 & \\
      & 0 & 1 & 2 & 0 & \\
      & 0 & 2 & 1 & 0 & \\
      & 1 & 0 & 0 & 2 & \\
      & 2 & 0 & 0 & 1 & \\
      & 0 & 1 & 1 & 1 & \\
      & 1 & 0 & 1 & 1 & \\
      & 1 & 1 & 0 & 1 & \\
      & 1 & 1 & 1 & 0 & \\
      \cmidrule(lr){2-6}
      & 3 & 0 & 0 & 0 & \multirow{2}{*}{$\frac{2}{9} \abs\alpha^6 \tau^3$} \\
      & 0 & 0 & 3 & 0 & \\
      \cmidrule(lr){2-6}
      & 0 & 3 & 0 & 0 & \multirow{2}{*}{$\frac{2}{9} \abs\beta^6 \tau^3$} \\
      & 0 & 0 & 0 & 3 & \\
    \bottomrule
    \end{tabular}
    \caption{Particle-number output distribution $f_3$ of the circuit in Fig.~\hyperref[fig:Implementation]{1c)} for a lossy NOON state ($N=1$) over all possible particle numbers $N_2^A$, $N_2^B, N_3^A, N_3^B$ grouped by total particle number $N_{\text{tot}} = N_2^A + N_2^B + N_3^A + N_3^B$.}
    \label{tab:PhotonNumberDistributionsp3Modified}
\end{table}

\clearpage

\bibliography{references.bib}

\end{document}